\documentclass[prd,preprint,showpacs,preprintnumbers,nofootinbib,eqsecnum,superscriptaddress]{revtex4-1}

 \usepackage[dvips,final]{graphicx}
  \usepackage{amssymb}
   \usepackage{amsmath}
    \usepackage{amsfonts}
     \usepackage{epsfig}
      \usepackage{bm}% bold math

\usepackage{mathpazo}

\usepackage[section]{placeins}

\usepackage{multirow}
\usepackage{ctable}
\usepackage{booktabs}
\usepackage{array}
\usepackage{tabularx}
\usepackage{xcolor}
\usepackage{pstricks}
\begin{document}

%\vfill
\title{Single-diffractive production of dijets\\
within the \bm{$k_t$}-factorization approach}

\author{Marta {\L}uszczak}
\email{luszczak@ur.edu.pl} 
\affiliation{Faculty of Mathematics and Natural Sciences, University of Rzesz\'ow, PL-35-959 Rzesz\'ow, Poland}

\author{Rafa{\l} Maciu{\l}a}
\email{rafal.maciula@ifj.edu.pl} 
\affiliation{Institute of Nuclear Physics PAN, PL-31-342 Cracow, Poland}

\author{Antoni Szczurek\footnote{also at University of Rzesz\'ow, PL-35-959 Rzesz\'ow, Poland}}
\email{antoni.szczurek@ifj.edu.pl}
\affiliation{Institute of Nuclear Physics PAN, PL-31-342 Cracow, Poland}

\author{Izabela Babiarz}
\email{i.babiarz@gmail.com} 
\affiliation{Faculty of Mathematics and Natural Sciences, University of Rzesz\'ow, PL-35-959 Rzesz\'ow, Poland}

\date{\today}

\begin{abstract}
We discuss single-diffractive production of dijets. The cross section is
calculated within the resolved pomeron picture, for the first time in 
the $k_t$-factorization approach, neglecting transverse momentum of the pomeron.
We use Kimber-Martin-Ryskin unintegrated parton (gluon, quark,
antiquark) distributions (UPDF) both in the proton as well as in the
pomeron or subleading reggeon.  The UPDFs are calculated based on 
conventional MMHT2014nlo PDFs in the proton and  H1 collaboration 
diffractive PDFs used previously in the analysis of diffractive 
structure function and dijets at HERA.  
For comparison we present results of calculations performed 
within collinear-factorization approach.
Our results remaind those obtained in the NLO approach.
The calculation is (must be) supplemented by the so-called gap survival 
factor which may, in general, depend on kinematical variables. We try 
to describe the existing data from Tevatron
and make detailed predictions for possible LHC measurements. 
Several differential distributions are calculated. The  $\overline{E}_T$,
 $\overline{\eta}$ and $x_{\bar p}$ distributions are compared with the Tevatron data.
A reasonable agreement is obtained for the first two distributions. 
The last one requires to introduce a gap survival factor which depends 
on kinematical variables. We discuss how the phenomenological dependence
on one kinematical variable may influence dependence on other variables
such as ${\overline E}_T$ and ${\overline \eta}$. 
Several distributions for the LHC are shown.
\end{abstract}


\maketitle

%----------------------------
\section{Introduction}
%----------------------------
The hard diffractive processes are related  to the production of a system 
with large mass (gauge boson, Higgs boson), or large invariant mass (dijets), 
and a presence of a rapidity gap somewhere in rapidity space. Several hard diffractive processes were studied in the past.
The gap may be in diffrent places with respect to final state objects, 
e.g. between forwardly produced proton and a “hard” system (hard single 
diffractive process) or between jets (jet-gap-jet topology) or quarkonia 
(quarkonium-gap-quarkonium).  Another category are exclusive diffractive
processes (Higgs, dijets, $\gamma \gamma$, pair of heavy quarks $Q \bar Q$, etc.)
Several other processes are possible in general, many of them not studied so far.

In the present paper we discuss  single-diffractive production of
dijets. This process was discussed in the past for photo- and
electro-production \cite{Klasen:2004qr,Kaidalov:2009fp,Klasen:2010vk,
Guzey:2016awf} as well as for
proton-proton or proton-antiproton collisions \cite{Appleby:2001xk,Kaidalov:2003gy,Klasen:2009bi,Royon:2013kla,Marquet:2013rja}.  The
hard single diffractive processes are treated usually in the resolved pomeron
picture with a pomeron being a virtual but composed (of partons)
object. This picture was used with a success for the description of 
hard diffractive processes studied extensively at HERA.
This picture was tried to be used also at hadronic collisions. 
A few processes were studied experimentally at the Tevatron 
\cite{Abe:1997rg,Affolder:1999hm,Affolder:2000hd,Affolder:2000vb,Affolder:2001zn,Wang:2000dq,Affolder:2001nc,Aaltonen:2007am,Aaltonen:2007hs,Aaltonen:2010qe,Aaltonen:2011hi} including the dijet production.

The related calculation were performed so far 
in the context of collinear-factorization approach. The corresponding 
parton distributions in pomeron, or equivalently so-called
diffractive parton distributions in the proton, were fitted  
so far to the HERA data. The distributions should be universal so, 
in principle, can be used in proton-proton collisions. In $p p$ or $p
\bar p$ collsions the strong nonperturbative interactions can easily 
destroy the rapidity gap associated with pomeron (or other
color-singlet) exchange. This effect is of nonperturbative nature 
and therefore difficult to be controlled.
There were several attempts to understand the related suppression 
of the hard diffractive cross sections.
Usually the effect is quantified by a phenomenological gap survival
factor. The factor is known to be energy dependent because the
nonperturbative soft interactions are known to be energy dependent.
In general, the survival probability may depend on other kinematical variables.
Recently the gap survival factor was studied for jet-gap-jet processes  
\cite{Babiarz:2017jxc} and the dependence on the gap sizes was discussed 
in the picture of multiple parton scattering. In our opinion we are still far
from the full understanding of the dynamical effect. 

In the present paper we intend to treat the single-diffractive dijet 
production for the first time within the $k_t$-factorization approach. 
Similar approach was used recently for the  single-diffractive
production of $c \bar c$ pairs \cite{Luszczak:2016csq}.
The $k_t$-factorization approach was also used recently for 
non-diffractive dijet \cite{Nefedov:2013ywa}, three- \cite{vanHameren:2013fla} or even four-jet 
production \cite{Kutak:2016mik}. In particular, we wish to compare results 
obtained within collinear-factorization and $k_t$-factorization
approaches. A comparison with the Tevatron data is planned. We would like 
to make also predictions for the LHC.

%----------------------------------------
\section{A sketch of the approach}
%-----------------------------------------

In this paper we follow the theoretical framework proposed very recently by three of us in Ref.~\cite{Luszczak:2016csq}. There, some new ideas for calculation of diffractive cross sections were put forward and applied in the case of single-diffractive production of charm at the LHC.
According to this approach, the standard resolved pomeron model \cite{IS}, usually based on the leading-order (LO) collinear approximation, is extended by adopting a framework of the $k_{t}$-factorization as an effective way to include higher-order corrections. It was shown several times, that the $k_{t}$-factorization approach is very usefull in this context and especially efficient in the studies of kinematical correlations (see \textit{e.g.} Refs.~\cite{Maciula:2013wg,Nefedov:2013ywa}).

%-----------------------------------------------------------------------------
\begin{figure}[!htbp]
\begin{minipage}{0.45\textwidth}
 \centerline{\includegraphics[width=1.0\textwidth]{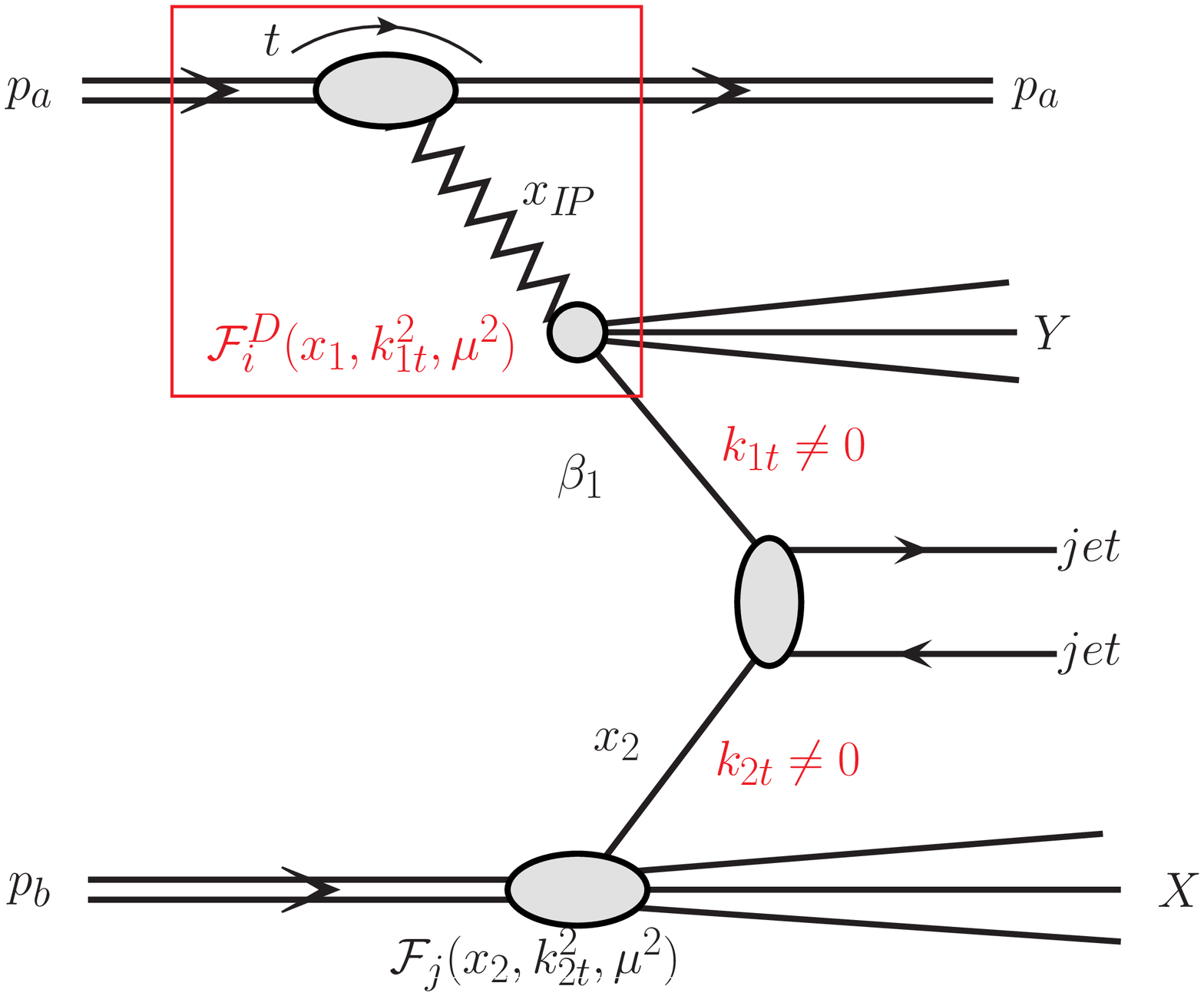}}
\end{minipage}
%\hspace{0.5cm}
\begin{minipage}{0.45\textwidth}
 \centerline{\includegraphics[width=1.0\textwidth]{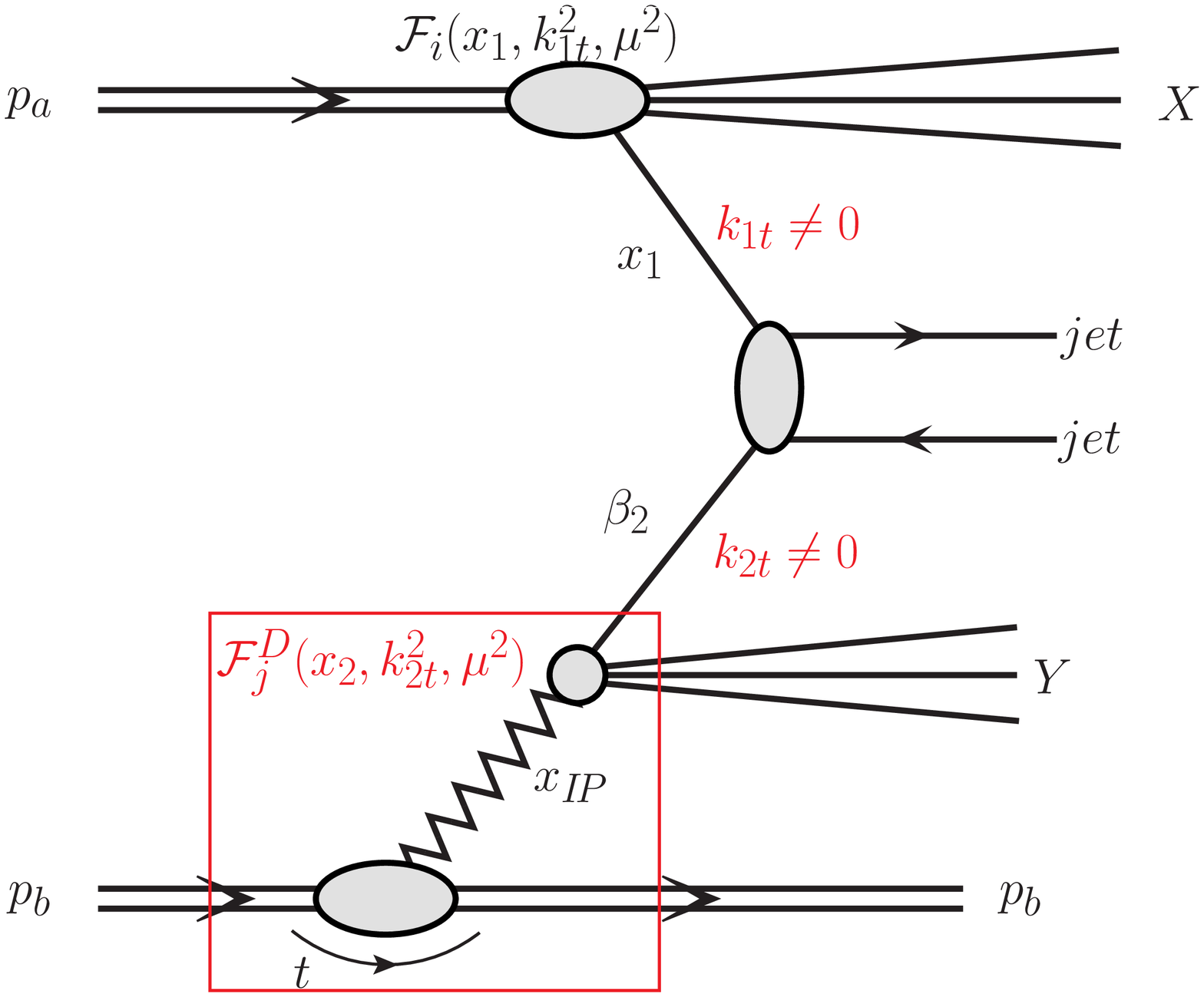}}
\end{minipage}
\caption{
\small A diagrammatic representation of the considered mechanisms of single-diffractive dijet production within resolved pomeron model extended in the present paper to the $k_{t}$-factorization approach.
}
 \label{fig:mechanism}
\end{figure}
%------------------------------------------------------------------------------

A sketch of the mechanisms under consideration, relevant for the inclusive single-diffractive production of dijets in $pp$ or $p\bar p$ collisions, with the notation of kinematical variables and with some theoretical ingredients used in the following is shown in Fig.~\ref{fig:mechanism}.

According to the approach introduced above, the cross section for inclusive single-diffractive production of dijet, for both considered diagrams (left and right panel of Fig.~\ref{fig:mechanism}), can be written as:
\begin{eqnarray}
d \sigma^{SD(1)}({p_{a} p_{b} \to p_{a} \; \mathrm{dijet} \; X Y}) &=& \sum_{i,j,k,l}
\int d x_1 \frac{d^2 k_{1t}}{\pi} d x_2 \frac{d^2 k_{2t}}{\pi} \; d {\hat \sigma}({i^{*}j^{*} \to kl }) \nonumber \\
&& \times \; {\cal F}_{i}^{D}(x_1,k_{1t}^2,\mu^2) \cdot {\cal F}_{j}(x_2,k_{2t}^2,\mu^2) ,
\label{SDA_formula}
\end{eqnarray}
\begin{eqnarray}
d \sigma^{SD(2)}({p_{a} p_{b} \to \mathrm{dijet} \; p_{b} \; X Y}) &=& \sum_{i,j,k,l}
\int d x_1 \frac{d^2 k_{1t}}{\pi} d x_2 \frac{d^2 k_{2t}}{\pi} \; d {\hat \sigma}({i^{*}j^{*} \to kl }) \nonumber \\
&& \times \; {\cal F}_{i}(x_1,k_{1t}^2,\mu^2) \cdot {\cal F}_{j}^{D}(x_2,k_{2t}^2,\mu^2),
\label{SDB_formula}
\end{eqnarray}
where ${\cal F}_{i}(x,k_{t}^2,\mu^2)$ are the "conventional" unintegrated ($k_{t}$-dependent) parton distributions (UPDFs) in the proton and ${\cal F}_{i}^{D}(x,k_{t}^2,\mu^2)$ are their diffractive counterparts -- which we will call here diffractive UPDFs (dUPDFs). The latter can be interpreted as a probability of finding a parton \textit{i} with longitudinal momentum fraction $x$ and transverse momentum (virtuality) $k_{t}$ at the factorization scale $\mu^{2}$ assuming that the proton which lost a momentum fraction $x_{I\!P}$ remains intact.

The $2 \to 2$ partonic cross sections in Eqs.(\ref{SDA_formula}) and (\ref{SDB_formula}) read: 
\begin{eqnarray}
d {\hat \sigma}({i^{*}j^{*} \to kl}) &=& \frac{d^3 p_1}{2 E_1 (2 \pi)^3} \frac{d^3 p_2}{2 E_2 (2 \pi)^3}
(2 \pi)^2 \delta^{2}(p_1 + p_2 - k_1 - k_2) \times \overline{|{\cal M}_{i^* j^* \to kl}(k_{1},k_{2})|^2} \; \nonumber \\
\label{elementary_cs}
\end{eqnarray}
with $i,j,k,l = g, u, d, s, \bar{u}, \bar{d}, \bar{s}$, where $p_1, E_1$ and $p_2, E_2$ are the momenta and energies of outgoing partons, respectively, and ${\cal M}_{i^* j^* \to kl}(k_{1},k_{2})$ are the off-shell matrix elements for the $i^* j^* \to kl$ subprocesses with initial state partons $i$ and $j$ being off mass shell.
In the numerical calculations here we include all $2 \to 2$ partonic channels:
\begin{eqnarray}
\#1 &=& g^{*} g^{*} \rightarrow g g    \, , \quad \#4 = g^{*} g^{*} \rightarrow q \bar{q}        \, , \quad \! \#7 = q^{*} \bar{q}^{*} \rightarrow g g  \nonumber \; , \\
\#2 &=& q^{*} g^{*} \rightarrow q g    \, , \quad \#5 = q^{*} \bar{q}^{*} \rightarrow q \bar{q}  \, , \quad \#8 = q^{*} q^{*} \rightarrow q q        \nonumber \; , \\
\#3 &=& g^{*} q^{*} \rightarrow g q    \, , \quad \#6 = q^{*} \bar{q}^{*} \rightarrow q'\bar{q}' , \quad \!\! \#9 = q^{*} q'^{*}\rightarrow q q'       \nonumber \; .
\end{eqnarray}
The relevant gauge-invariant off-shell matrix elements for each of the channels above can be calculated \textit{e.g.} within the method of parton reggeization. It was done recently in Ref.~\cite{Nefedov:2013ywa} where the matrix elements were presented in a very useful analytical form. 

As we have proposed very recently in Ref.~\cite{Luszczak:2016csq}, the diffractive UPDFs can be calculated using their collinear counterparts via the Kimber-Martin-Ryskin (KMR) method \cite{Kimber:2001sc,Watt:2003vf}. Then, the diffractive unintegrated parton distributions for gluon and quark are given by the following formulas:
\begin{eqnarray} \label{eq:UPDF}
  f^{D}_{g}(x,k_t^2,\mu^2) &\equiv& \frac{\partial}{\partial \log k_t^2}\left[\,g^{D}(x,k_t^2)\,T_g(k_t^2,\mu^2)\,\right] =
  T_g(k_t^2,\mu^2)\,\frac{\alpha_S(k_t^2)}{2\pi}\, \times  \\  
&& \!\!\!\int_x^1\! d z \left[\sum_q P_{gq}(z)\frac{x}{z}\;q^{D}\!\left(\frac{x}{z},k_t^2\right) + P_{gg}(z)\frac{x}{z}\;g^{D}\!\left(\frac{x}{z},k_t^2\right)\Theta\left(\Delta - z\right)\right], \nonumber
\end{eqnarray}
\begin{eqnarray} \label{eq:UPDF}
  f^{D}_{q}(x,k_t^2,\mu^2) &\equiv& \frac{\partial}{\partial \log k_t^2}\left[\,q^{D}(x,k_t^2)\,T_q(k_t^2,\mu^2)\,\right] =
  T_q(k_t^2,\mu^2)\,\frac{\alpha_S(k_t^2)}{2\pi}\, \times  \\  
&& \!\!\!\int_x^1\! d z \left[ P_{qq}(z)\frac{x}{z}\;q^{D}\!\left(\frac{x}{z},k_t^2\right)\Theta\left(\Delta - z\right) + P_{qg}(z)\frac{x}{z}\;g^{D}\!\left(\frac{x}{z},k_t^2\right)\right], \nonumber
\end{eqnarray}
where $g^{D}$ and $q^{D}$ are the collinear diffractive PDFs in the proton. The $P_{qq}, P_{qg}, P_{gq}$ and $P_{gg}$ are the usual unregulated LO DGLAP splitting functions and $T_{g}$ and $T_{q}$ are the gluon and quark Sudakov form factors, respectively. More details of the whole procedure and discussion of all of the ingredients can be found \textit{e.g.} in Ref.~\cite{Watt:2003vf}.  

According to the so-called proton-vertex-factorization, the diffractive collinear PDF in the proton, \textit{e.g.} for gluon, has the following generic form:
\begin{eqnarray}
g^D(x,\mu^2) = \int d x_{I\!P} d\beta \, \delta(x-x_{I\!P} \beta) 
g_{I\!P} (\beta,\mu^2) \, f_{I\!P}(x_{I\!P}) \, 
= \int_x^{x^{max}} {d x_{I\!P} \over x_{I\!P}} \, f_{I\!P}(x_{I\!P})  
g_{I\!P}({x \over x_{I\!P}}, \mu^2) , \nonumber \\  
\end{eqnarray} 
where $\beta = \frac{x}{x_{I\!P}}$ is the longitudinal momentum fraction of the pomeron carried by gluon and the flux of pomerons may be taken as:
\begin{equation}
f_{I\!P}(x_{I\!P}) = \int_{t_{min}}^{t_{max}} dt \, f(x_{I\!P},t).
\end{equation}
An analogous expression can be also written for the collinear diffractive quark distribution.

In this paper, the diffractive KMR UPDFs are calculated from the "H1 2006 fit A" diffractive collinear PDFs \cite{H1}, that are only available at next-to-leading order (NLO). In the calculation of the conventional non-diffractive KMR UPDFs the  collinear MMHT2014nlo PDFs \cite{Harland-Lang:2014zoa} were used. In the perturbative part of calculations we take running coupling constant $\alpha_{S}(\mu_{R}^{2})$ and the renormalization and factorization scales equal to $\mu^{2} = \mu_{R}^{2} = \mu_{F}^{2} = \frac{p_{1t}^{2} + p_{2t}^{2}}{2}$, where $p_{1t}$ and $p_{2t}$ are the transverse momenta of the outgoing jets.
%-------------------
\section{Results}
%-------------------

In this section we shall show results of our calculations. 
We shall start from a trial of the description
of the Tevatron experimental data \cite{Affolder:2000vb,Affolder:2001zn}.

%-------------------------
\subsection{Tevatron cuts}
%--------------------------

\begin{figure}[!htbp]
\begin{minipage}{0.47\textwidth}
 \centerline{\includegraphics[width=1.0\textwidth]{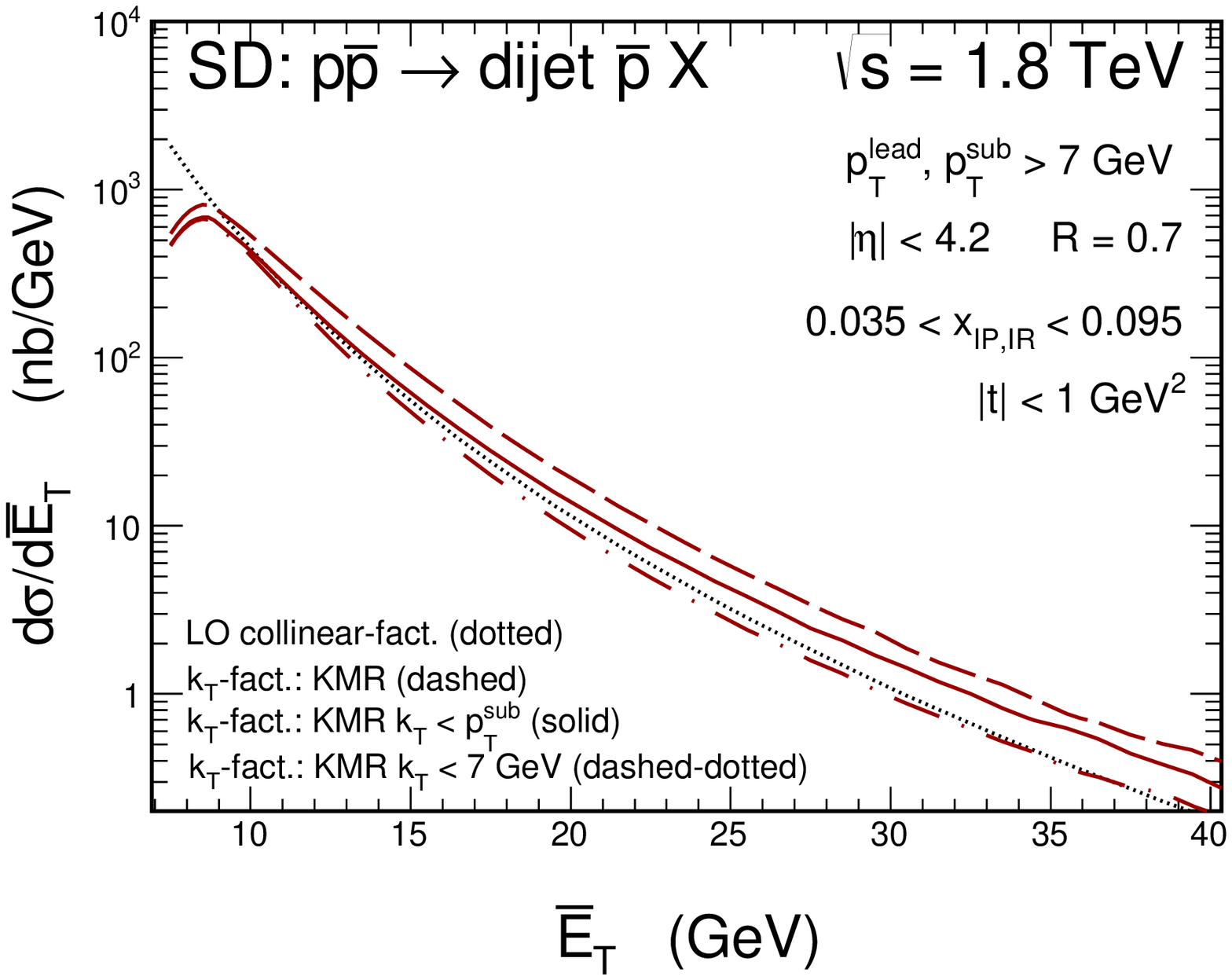}}
\end{minipage}
\hspace{0.5cm}
\begin{minipage}{0.47\textwidth}
 \centerline{\includegraphics[width=1.0\textwidth]{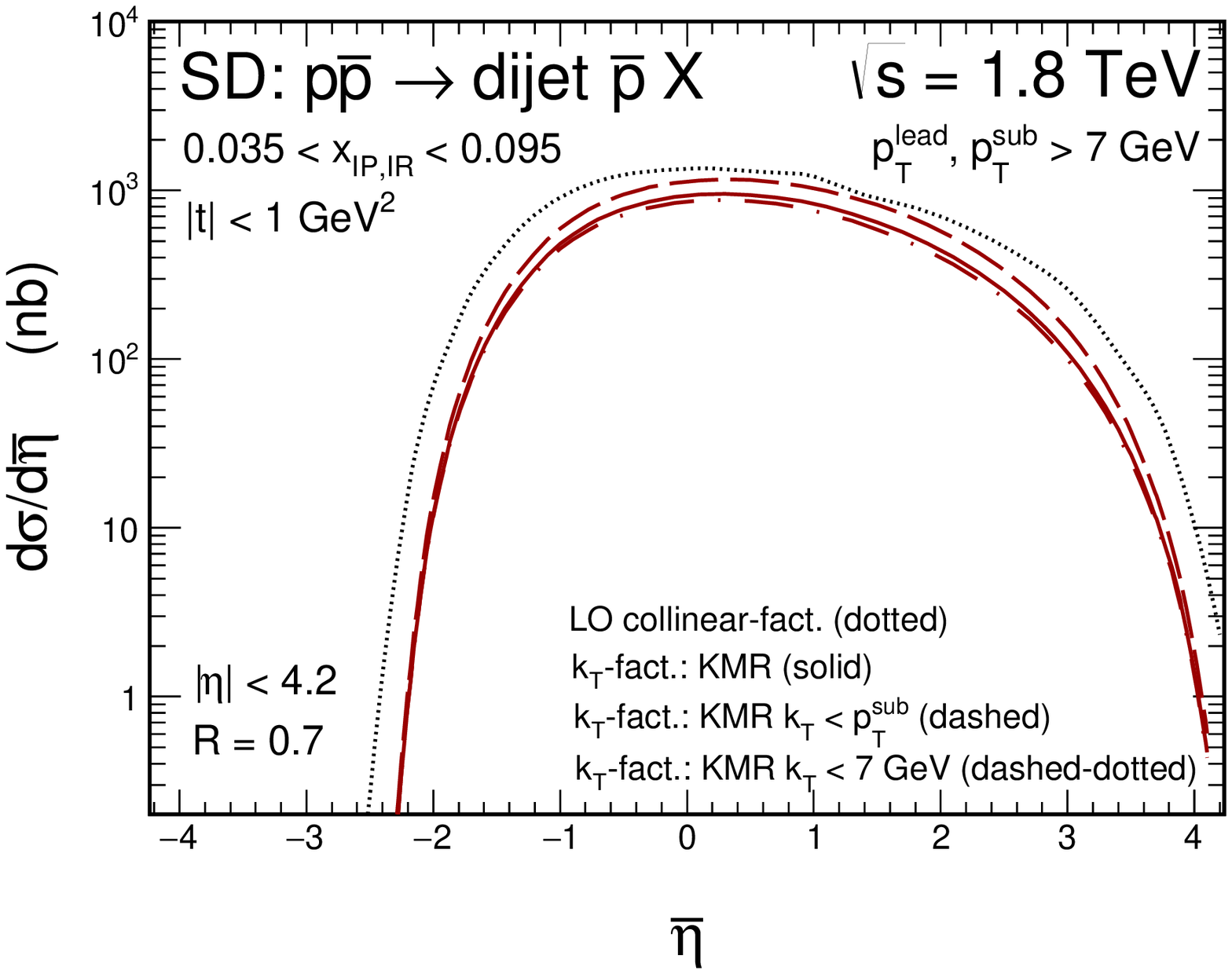}}
\end{minipage}
   \caption{
\small Distribution in average ${\overline E}_T$ (left panel) and
in average ${\overline \eta}$ (right panel). Here $S_G$ = 0.1.
}
 \label{fig:dsig-1800_SD}
\end{figure}

We start from showing our results for ${\overline E}_T = \frac{E_{1T} + E_{2T}}{2}$
and ${\overline \eta} = \frac{\eta_1 + \eta_2}{2}$ distributions, see Fig.~\ref{fig:dsig-1800_SD}.
In this calculation the pomeron/reggeon longitudinal momentum 
fraction was limited as in experimental case \cite{Affolder:2000vb,Affolder:2001zn} 
to 0.035 $< x_{I\!P,I\!R} <$ 0.095.
We show both naive result obtained with the KMR UGDF (dashed line)
as well as similar results with limitations on parton transverse
momenta $k_T < p_T^{sub}$ (solid line) and $k_T < 7 $ TeV (dash-dotted line).
The first limitation was proposed for standard nondiffractive jets
\cite{Nefedov:2013ywa}. The latter limitation is related to the lower experimental
cut on jet transverse momenta.
For comparison we show also distribution obtained in leading-order
collinear factorization approach (dotted line).
A large difference can be seen close to the lower transverse momentum
cut. Similar effect was discussed recently for four jet production
in \cite{Kutak:2016mik}.

%-----------------------------------------------------------------------------
%------------------------------------------------------------------------------

\begin{figure}[!htbp]
\begin{minipage}{0.38\textwidth}
 \centerline{\includegraphics[width=1.0\textwidth]{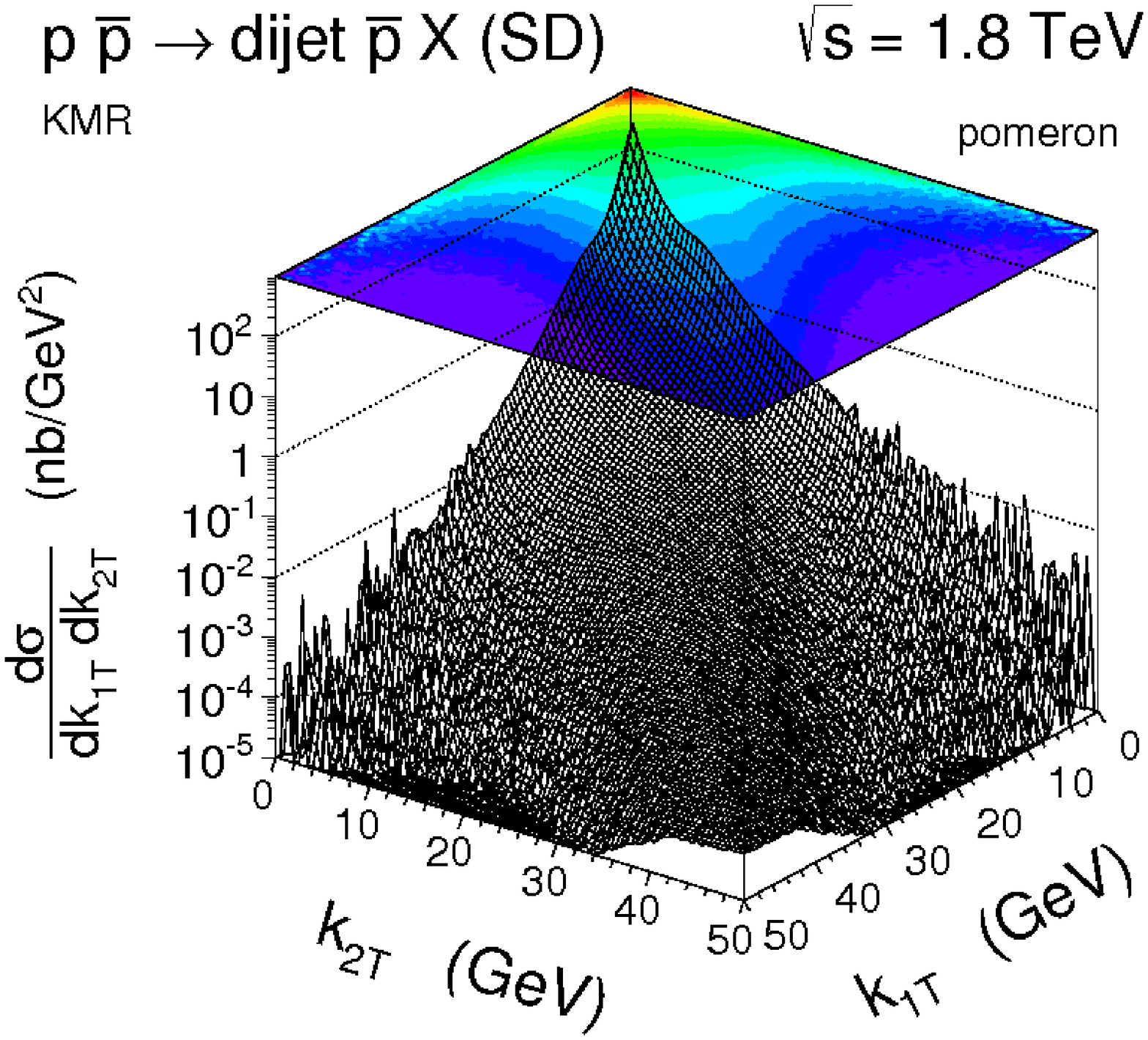}}
\end{minipage}
\hspace{0.5cm}
\begin{minipage}{0.38\textwidth}
 \centerline{\includegraphics[width=1.0\textwidth]{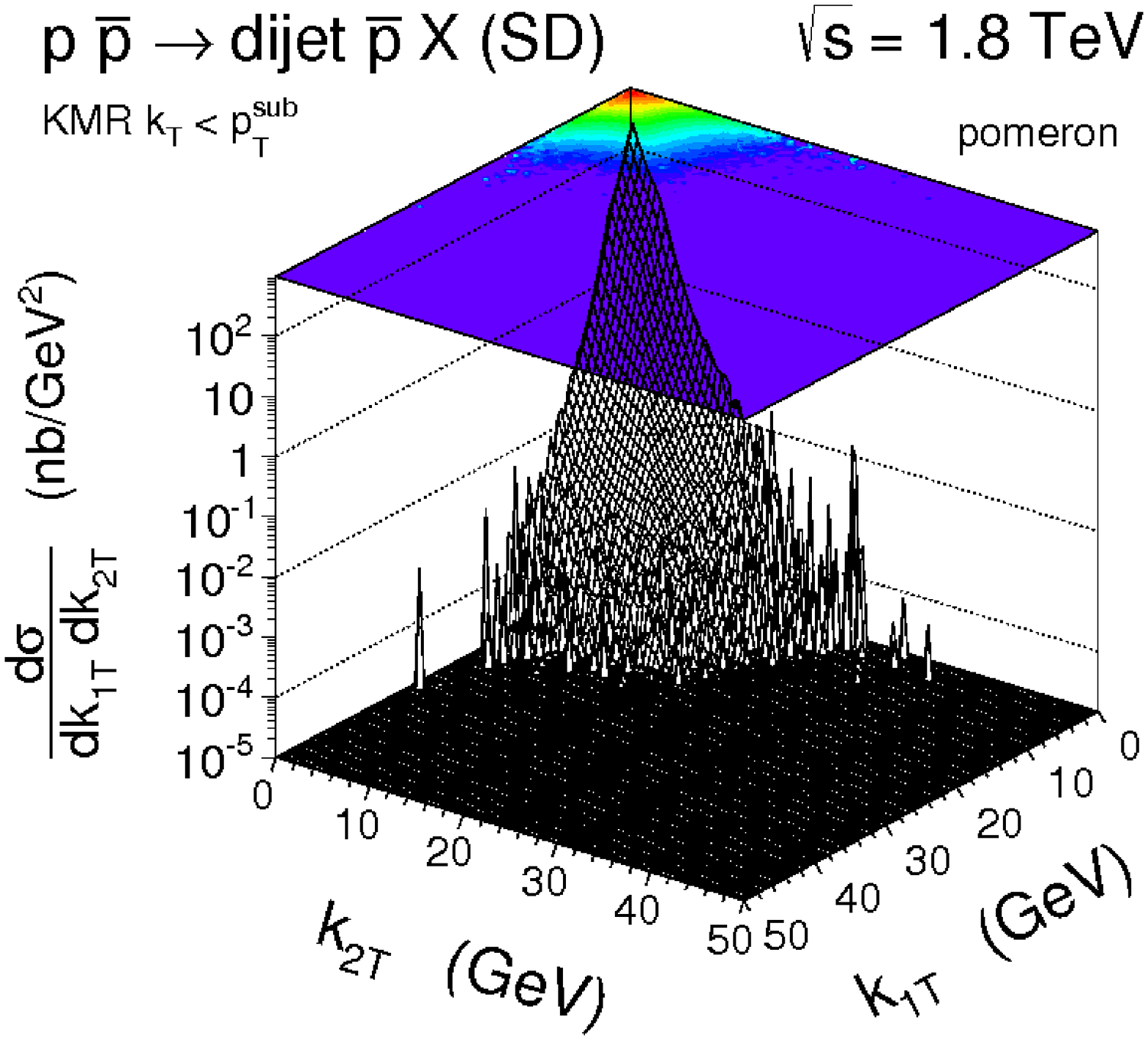}}
\end{minipage}
   \caption{
\small 
Two dimensional distribution in transverse momenta of partons on 
the nondiffractive side ($k_{1T}$) and on the diffractive side
($k_{2T}$). Here $S_G$ = 0.1.
}
 \label{fig:map_q1tq2t-1800}
\end{figure}

Fig.~\ref{fig:map_q1tq2t-1800} shows somewhat theoretical two-dimensional distribution in
transverse momenta of partons. Surprisingly the distribution 
is almost symmetric in $k_{1T}$ and $k_{2T}$.
The limitation on parton transverse momenta $k_T < p_T^{sub}$ 
makes the two-dimensional distribution much narrower, 
although the consequences on distribution in transverse
momenta and rapidity are not dramatic as has been already shown in Fig.~\ref{fig:dsig-1800_SD}.

%-----------------------------------------------------------------------------
\begin{figure}[!htbp]
\begin{minipage}{0.47\textwidth}
 \centerline{\includegraphics[width=1.0\textwidth]{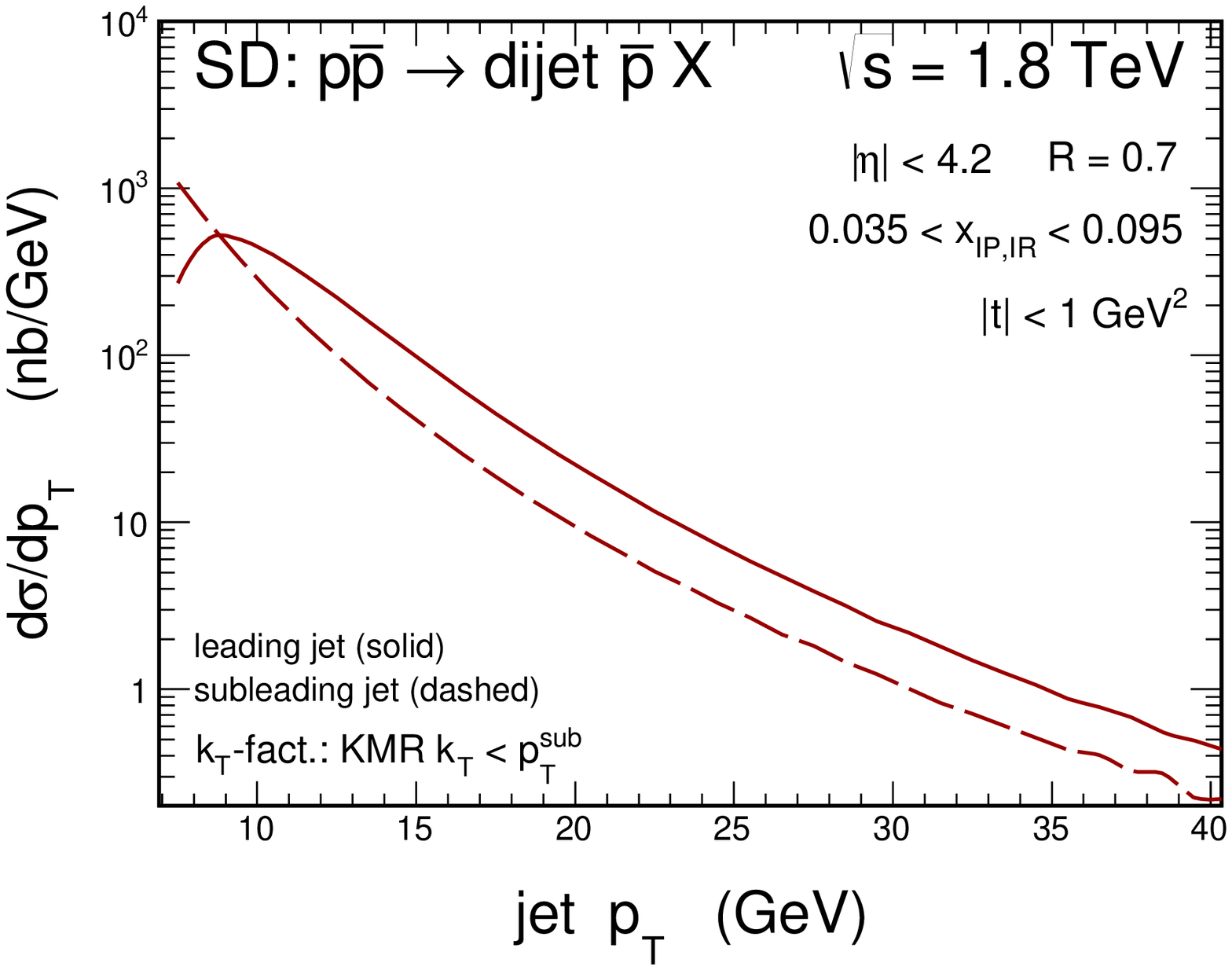}}
\end{minipage}
\hspace{0.5cm}
\begin{minipage}{0.47\textwidth}
 \centerline{\includegraphics[width=1.0\textwidth]{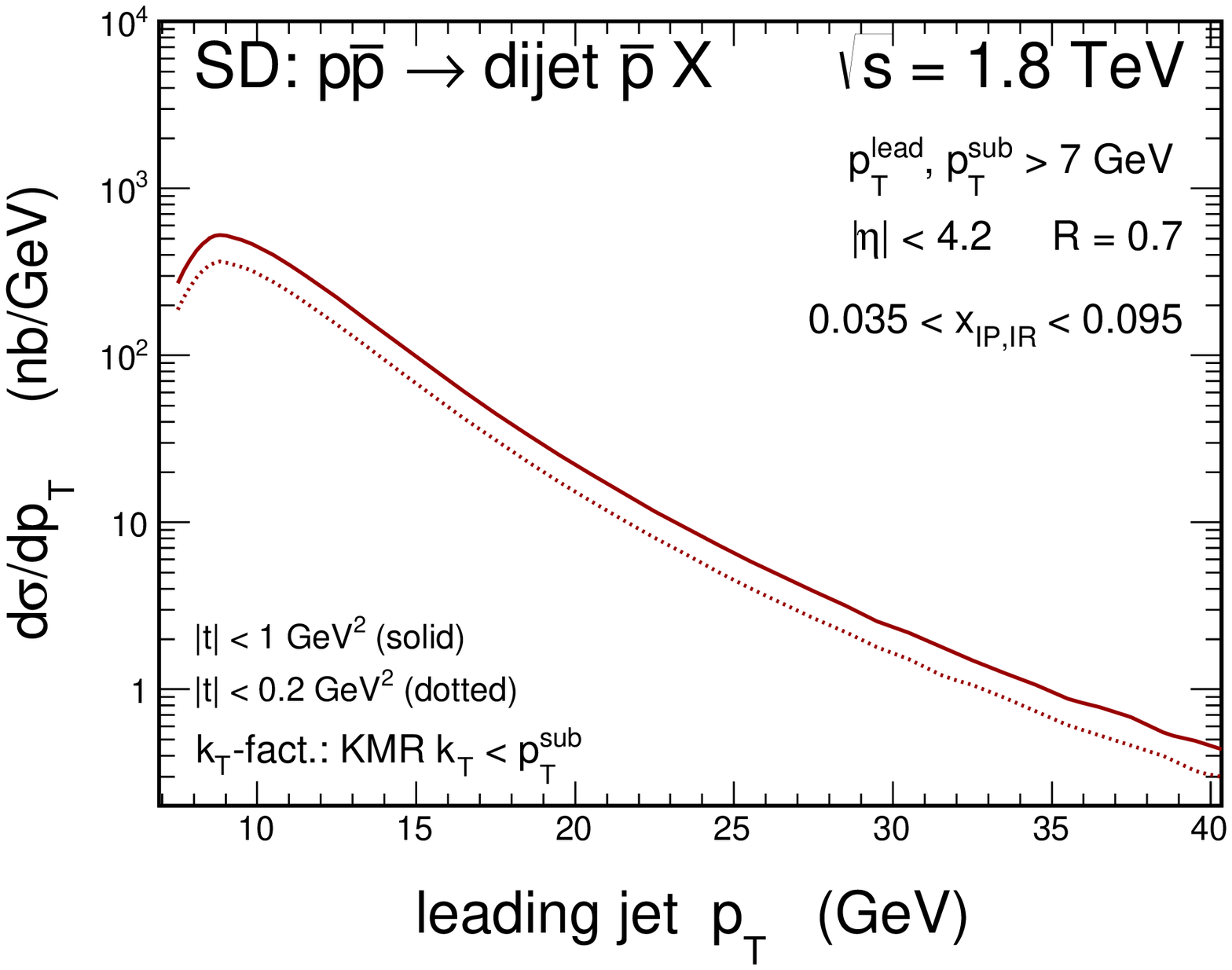}}
\end{minipage}
   \caption{
\small Transverse momentum distribution for leading and subleading jet (left
panel) and the influence of the cut on $t$ on the leading jet (right panel). Here $S_G$ = 0.1.
}
 \label{fig:dsig_dpt-1800}
\end{figure}

%------------------------------------------------------------------------------
In contrast to the leading-order collinear factorization approach,
in the $k_t$-factorization approach the transverse momentum distribution
of leading (solid) and subleading (dashed) jets differ as is shown in 
the left panel of Fig.~\ref{fig:dsig_dpt-1800} 

Here a standard cut on parton transverse momentum has been imposed.
The single diffractive cross section depends on the cut on
four-momentum squared transferred to the outgoing antiproton
(antiproton was measured in the CDF experiment).
The cut changes the cross section normalization but does not
modify the shape of the distribution. 

%-----------------------------------------------------------------------------
%------------------------------------------------------------------------------

\begin{figure}[!htbp]
\begin{minipage}{0.47\textwidth}
 \centerline{\includegraphics[width=1.0\textwidth]{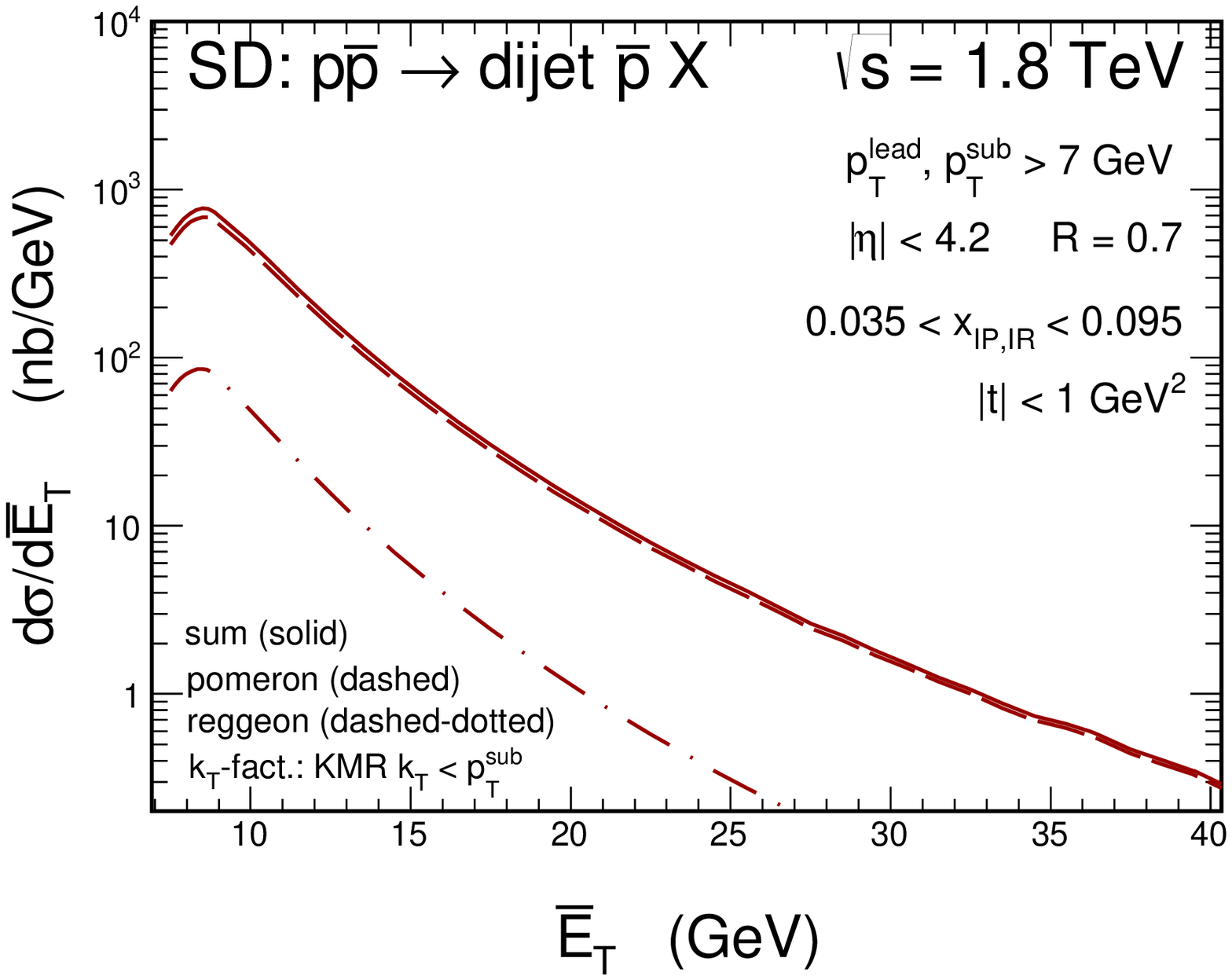}}
\end{minipage}
\hspace{0.5cm}
\begin{minipage}{0.47\textwidth}
 \centerline{\includegraphics[width=1.0\textwidth]{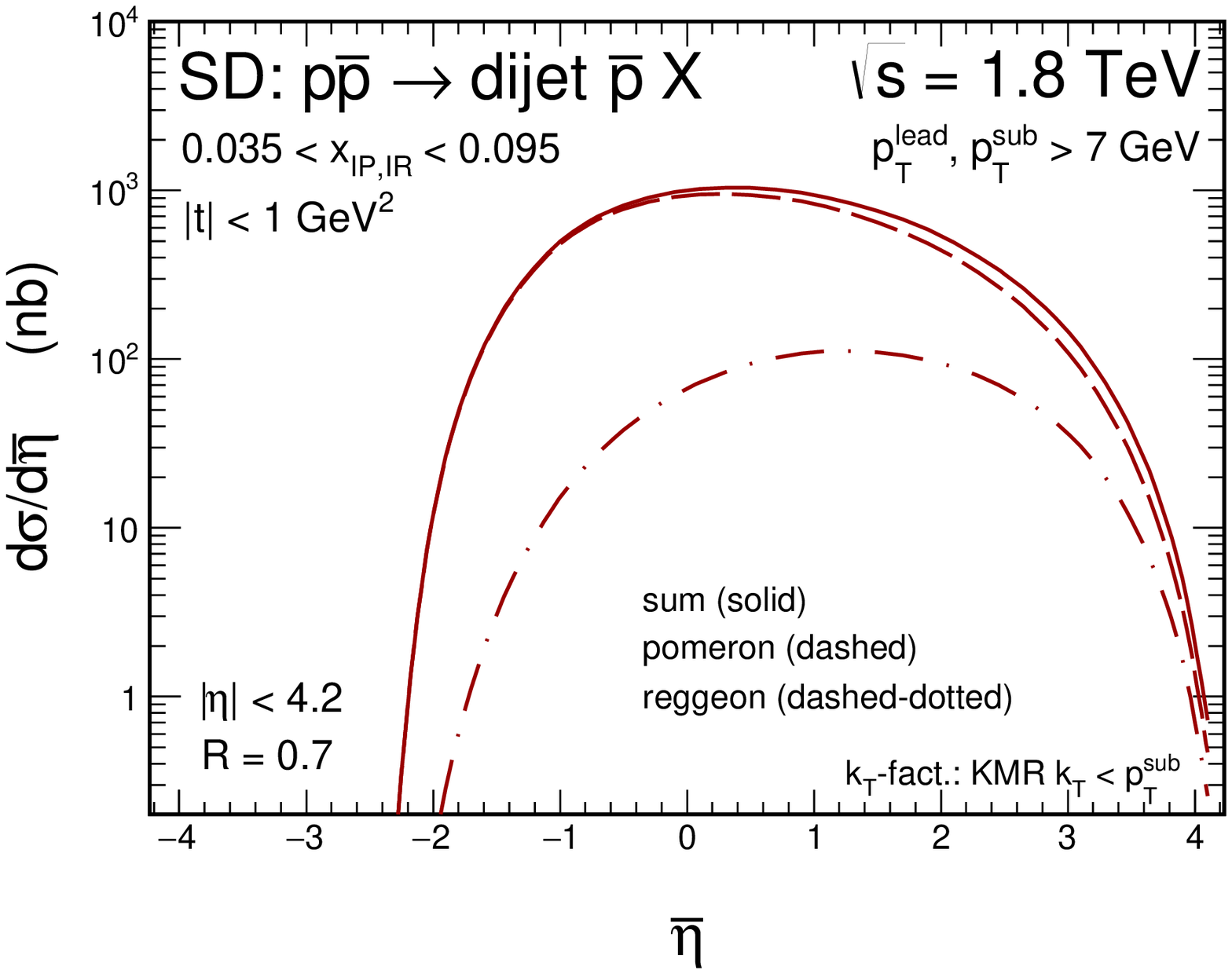}}
\end{minipage}
   \caption{
\small The pomeron and subleading reggeon contribution for
${\overline E}_T$ (left panel) and ${\overline \eta}$ (right panel).
}
 \label{fig:dsig_dpt-1800_R}
\end{figure}

In our calculation we include both pomeron and subleading reggeon
exchanges. In the selected range of $x_{I\!P}$ the pomeron contribution
is much bigger than the contribution of the subleading reggeon
as shown in Fig.~\ref{fig:dsig_dpt-1800_R}. The subleading reggeon contribution is about 10 \%
of the single diffractive cross section. For the average jet rapidity
distribution the situation is a bit more complicated.
Both contributions are of the same order for large ${\overline \eta}$.

%-----------------------------------------------------------------------------
%------------------------------------------------------------------------------

Now we would like to consider distributions that can be compared to the
experimental ones.

\begin{figure}[!htbp]
\begin{minipage}{0.47\textwidth}
 \centerline{\includegraphics[width=1.0\textwidth]{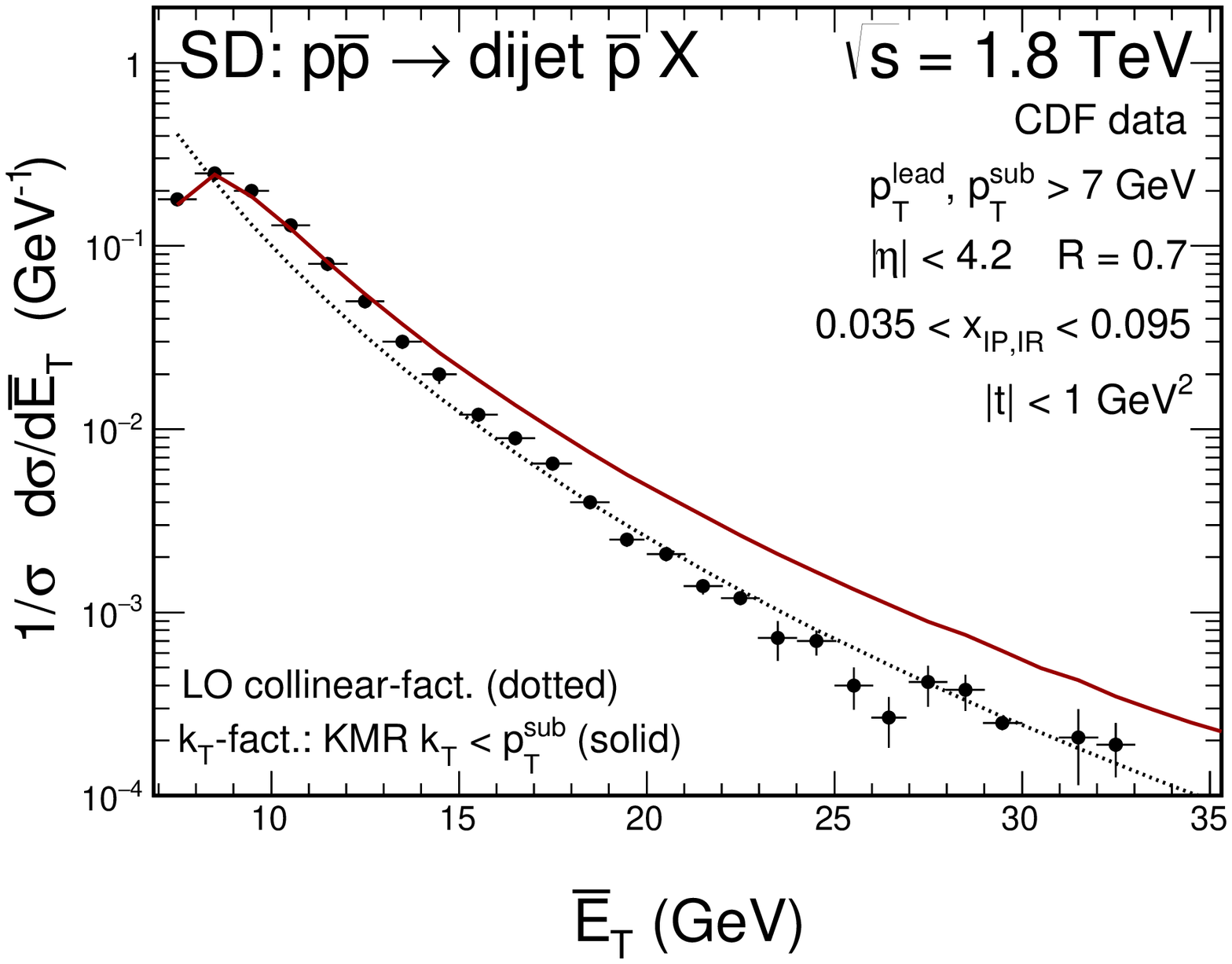}}
\end{minipage}
\hspace{0.5cm}
\begin{minipage}{0.47\textwidth}
 \centerline{\includegraphics[width=1.0\textwidth]{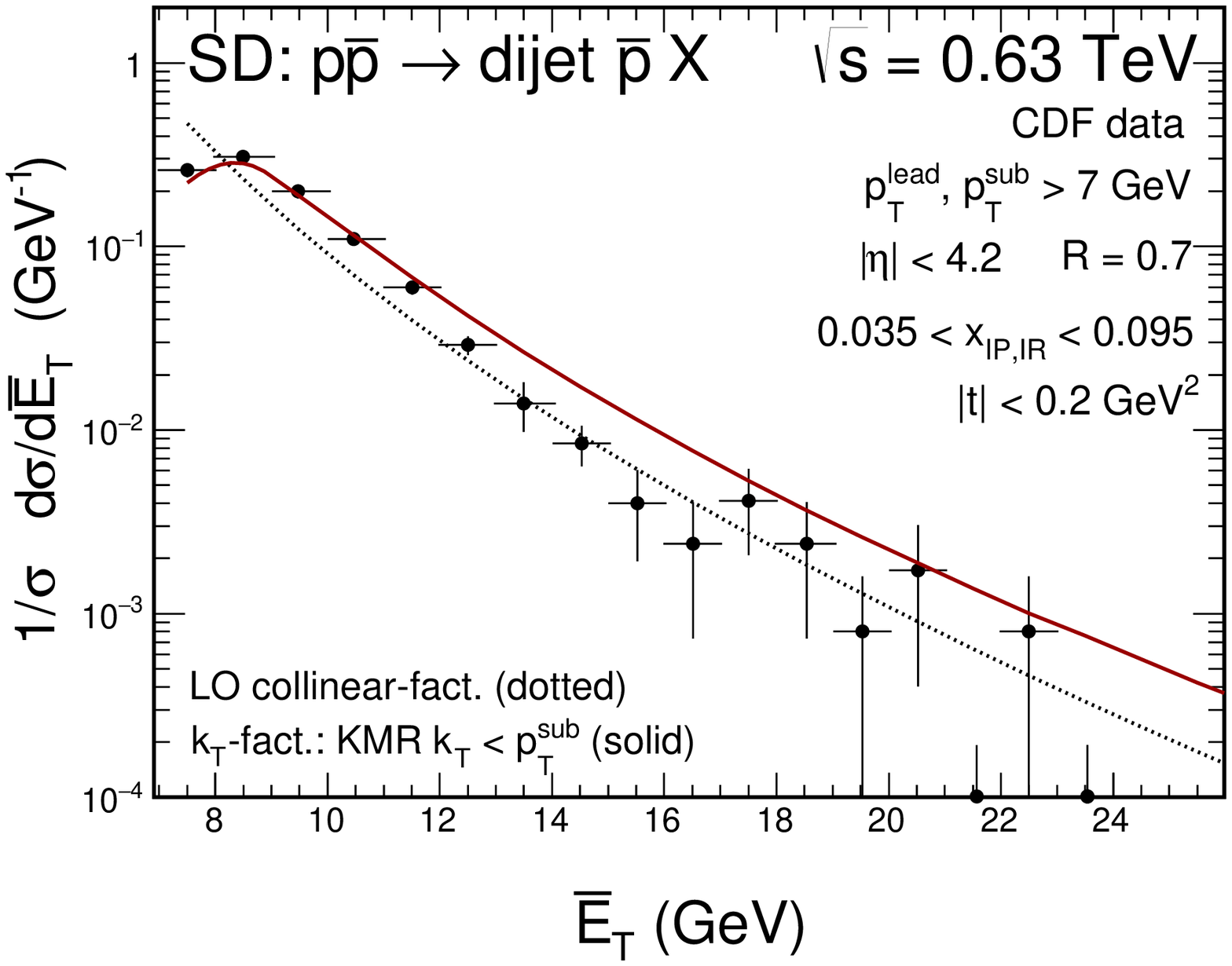}}
\end{minipage}
   \caption{
\small The average transverse energy distribution for $\sqrt{s}$ = 1.8 TeV
(left panel) and for $\sqrt{s}$ = 630 GeV (right panel).
%------------------------------------------------------------------
}
 \label{fig:cdf_pt}
\end{figure}

In Fig.~\ref{fig:cdf_pt} we show distribution in ${\overline E}_T$ for two collision
energies. While the $k_t$-factorization approach gives a better
description of the data close to the lower experimental cut on
jet transverse momenta, the collinear-factorization approach seems to
be better for larger transverse momenta. This is true for both Tevatron collision
energies. We do not have good understanding of the result.

%-----------------------------------------------------------------------------
%------------------------------------------------------------------------------

\begin{figure}[!htbp]
\begin{minipage}{0.47\textwidth}
 \centerline{\includegraphics[width=1.0\textwidth]{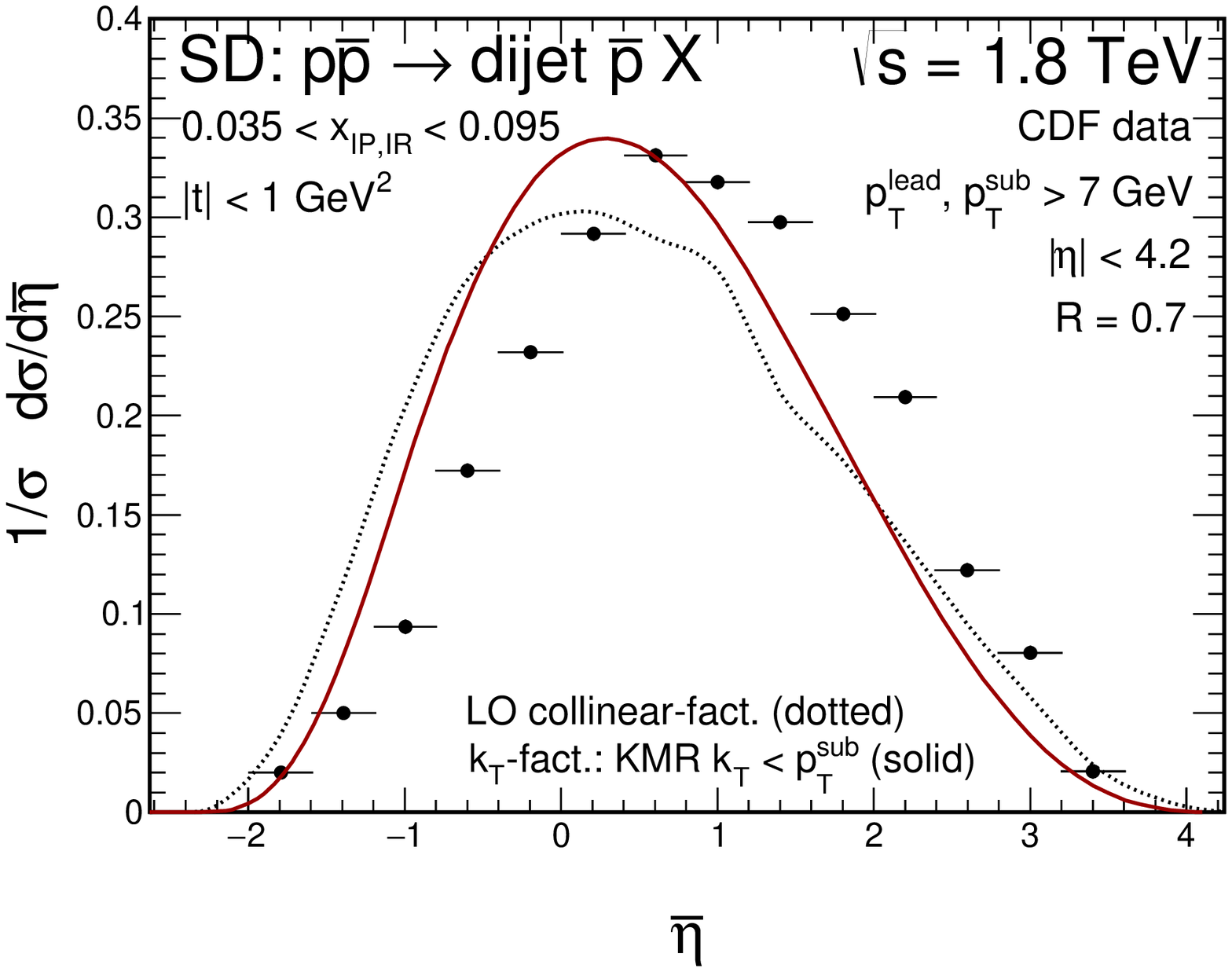}}
\end{minipage}
\hspace{0.5cm}
\begin{minipage}{0.47\textwidth}
 \centerline{\includegraphics[width=1.0\textwidth]{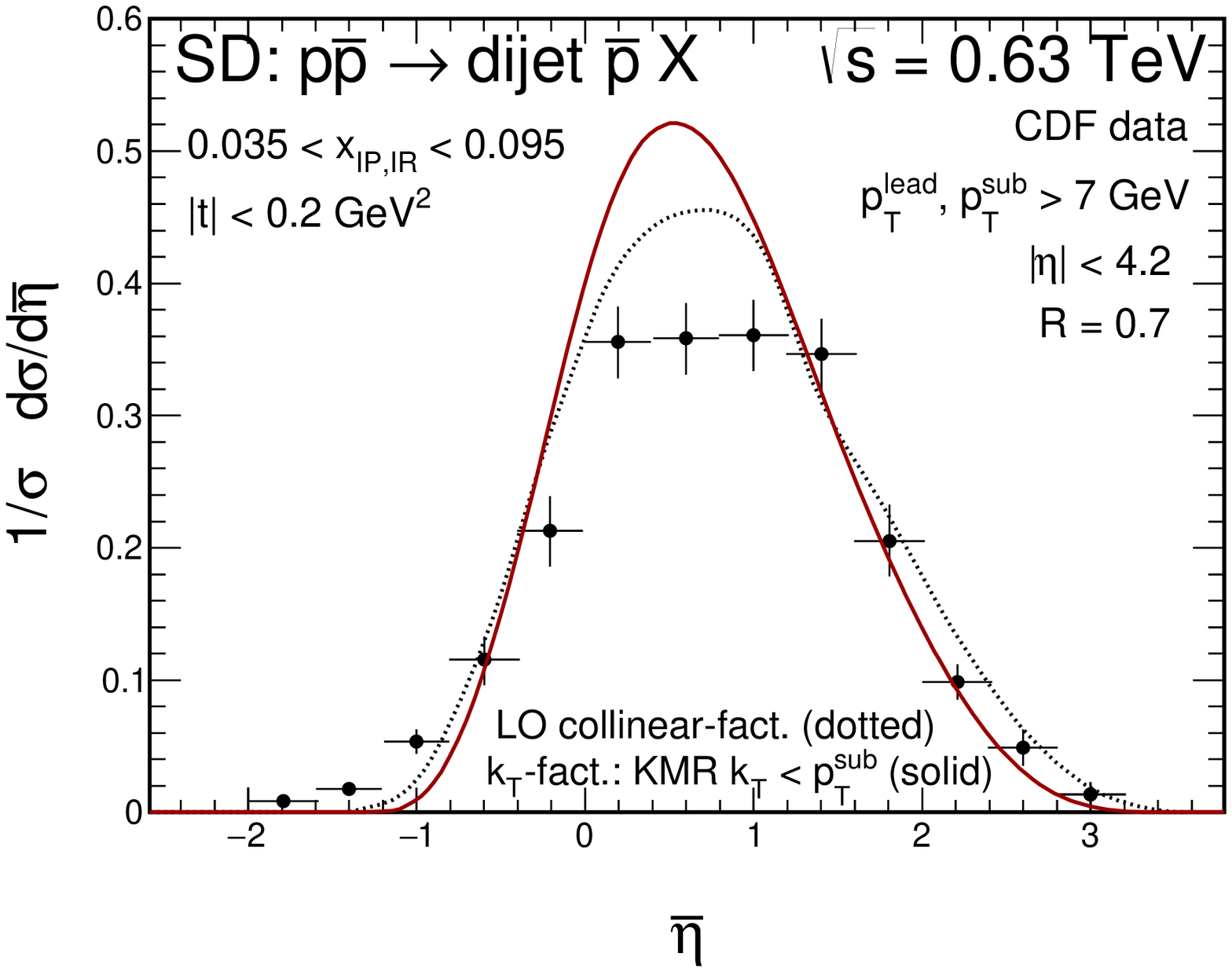}}
\end{minipage}
   \caption{
\small The average rapidity dsistribution for $\sqrt{s}$ = 1.8 TeV
(left panel) and for $\sqrt{s}$ = 630 GeV (right panel).
}
 \label{fig:cdf_y}
\end{figure}

In Fig.~\ref{fig:cdf_y} we show distributions in average jet rapidity again for 
the two collision energies. Here the $k_t$-factorization result 
better describes the experimental data than the result obtained in 
the collinear approach. The outgoing antiproton is at
$\eta \approx$ -6.05 for $\sqrt{s}$ = 1.8 TeV and $\eta \approx$ -5.53 for $\sqrt{s}$ = 630 GeV,
respectively.

%-----------------------------------------------------------------------------
\begin{figure}[!htbp]
\begin{minipage}{0.47\textwidth}
 \centerline{\includegraphics[width=1.0\textwidth]{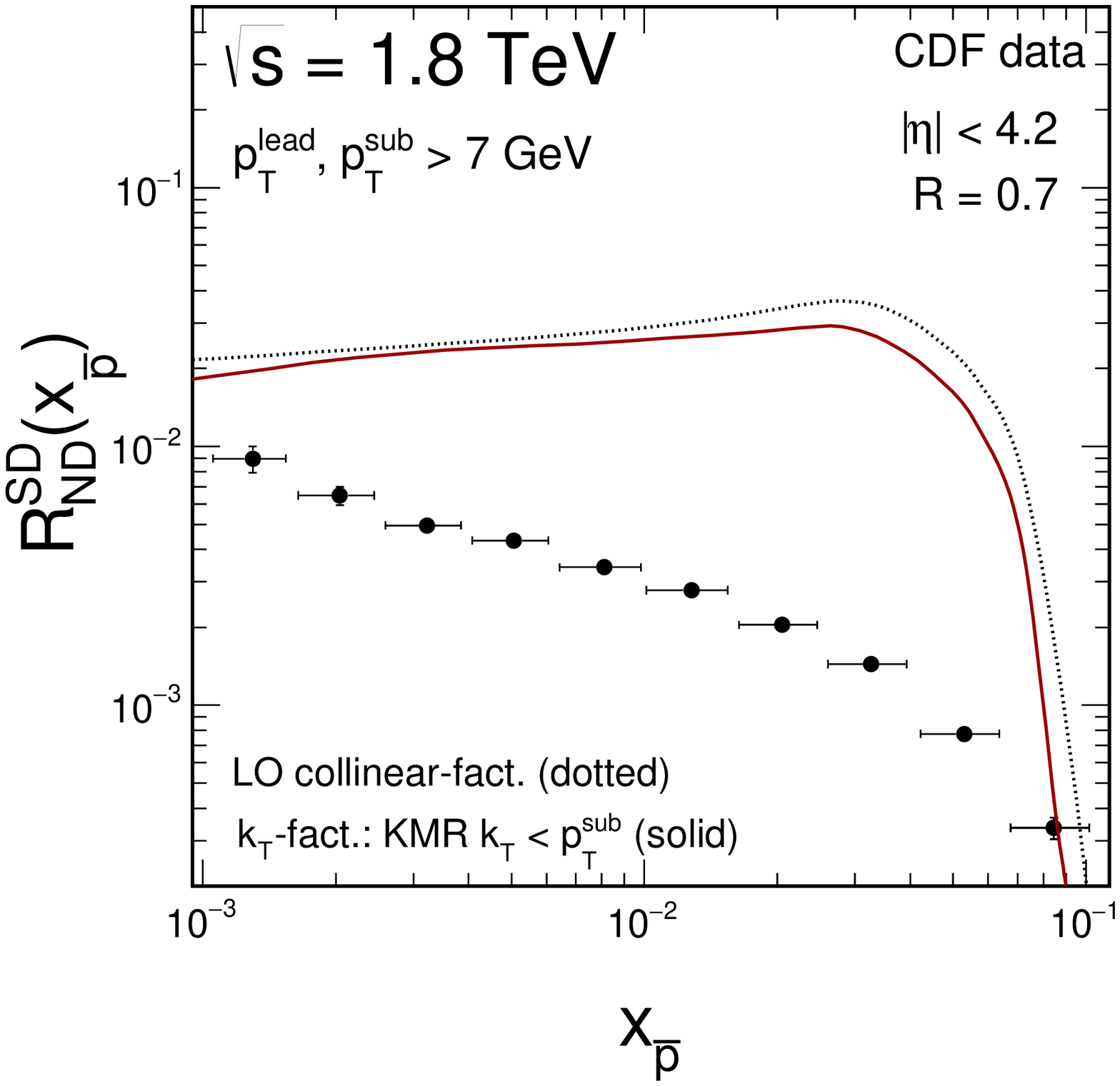}}
\end{minipage}
\hspace{0.5cm}
\begin{minipage}{0.47\textwidth}
 \centerline{\includegraphics[width=1.0\textwidth]{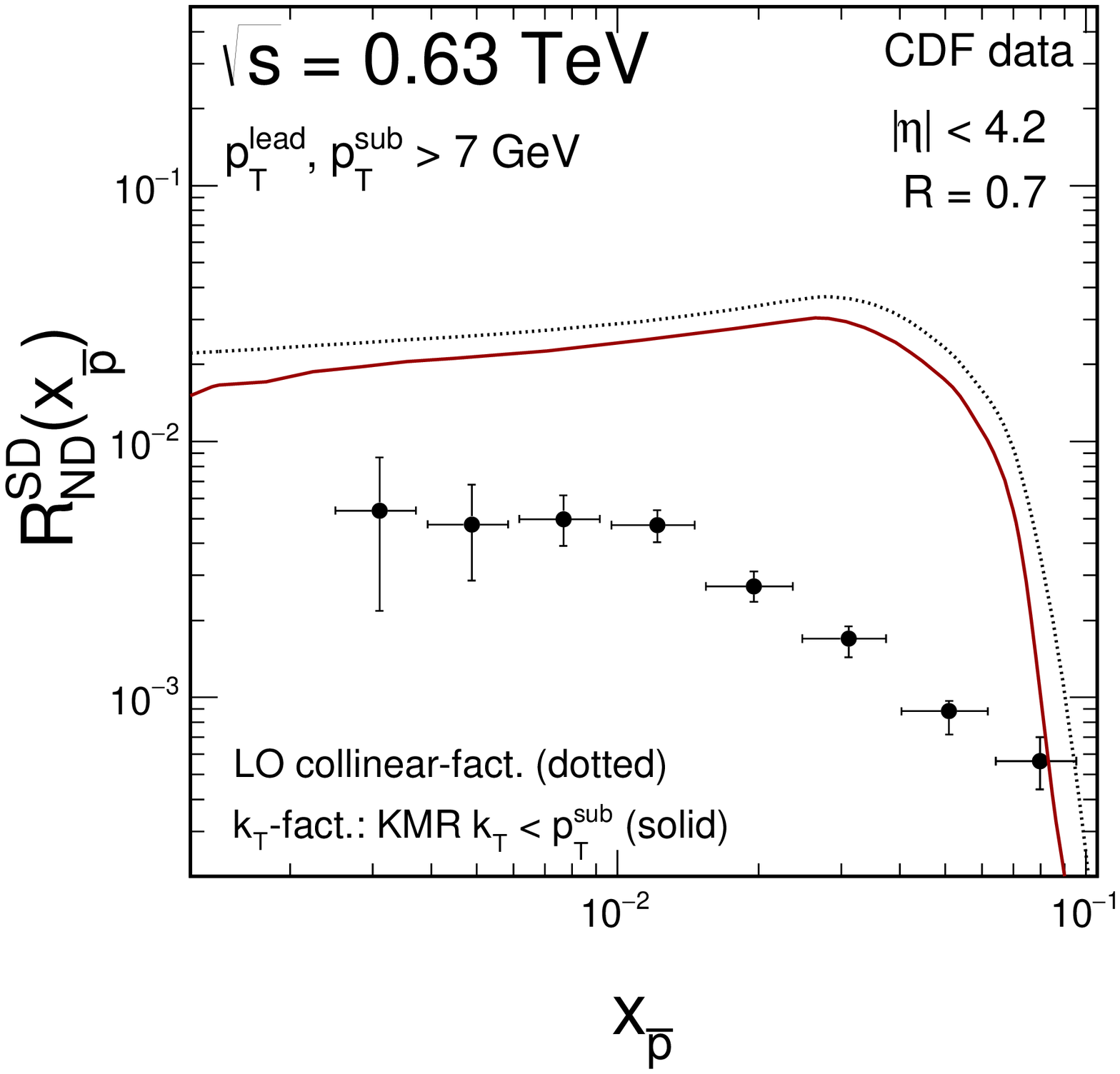}}
\end{minipage}
   \caption{
\small Distribution in $x_{\bar p}$ for $\sqrt{s}$ = 1.8 TeV
(left panel) and for $\sqrt{s}$ = 630 GeV (right panel). No gap survival factor was included here.
}
 \label{fig:ratio1}
\end{figure}

%------------------------------------------------------------------------------
Let us note here that both the experimental distributions
in ${\overline E}_T$ and in ${\overline \eta}$ are not absolutely
normalized (inspect description of "$y$" axes of Fig.~\ref{fig:cdf_pt} and Fig.~\ref{fig:cdf_y}).
On the theoretical side the absolute cross section depends on
gap survival factor which is not easy to calculated from first
principle. The CDF collaboration showed also distribution
in $x_{\bar p}$ normalized to the inclusive cross section.
Our theoretical result is clearly above the experimental result, see Fig.~\ref{fig:ratio1}.
Roughly a factor of order 0.1 is missing in our calculation
although the exact shape is not exactly the same.

%-----------------------------------------------------------------------------
%------------------------------------------------------------------------------

There can be several reasons of the disagreement of our results with
the CDF data. One of them is not a perfect extraction of the diffractive
distributions at HERA. Another one is the dependence of the gap survival factor 
on kinematical variables. This possibility will be discussed now
in the next subsection.

%---------------------------------------------------------
\subsection{Kinematical dependence of gap survival factor}
%---------------------------------------------------------

\begin{figure}[!htbp]
\begin{minipage}{0.47\textwidth}
 \centerline{\includegraphics[width=1.0\textwidth]{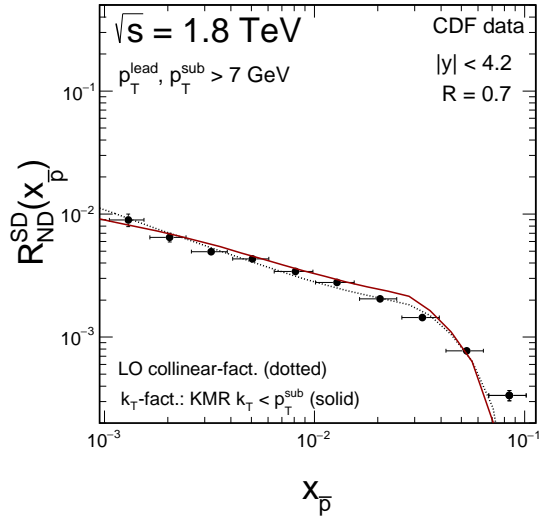}}
\end{minipage}
   \caption{
\small The ratio of single-diffractive to nondiffractive cross sections
as a function of $x_{\bar p}$. The lines are fits of $S_G$
to the CDF data.
}
 \label{fig:ratio2}
\end{figure}

In this section we assume that the gap survival factor is a function
of $x_{\bar p}$ only. This assumption is a bit academic but we
would like to see a possible influence of such a dependence on
other distributions. In Fig.~\ref{fig:ratio2} we show a fit to the data assuming
some functional form for $S_G(x_{\bar p})$,
$S_G(x_{\bar p}) = 0.0056*(x_{1}^{-0.6} + x_{1}^{-0.02})$ for collinear case and $S_G(x_{\bar p}) = 0.004*(x_{1}^{-0.6} + x_{1}^{-0.03})$ for $k_t$-factorization approach.
Our fit nicely describes the CDF data at $\sqrt{s}$ = 1.8 TeV.

%-----------------------------------------------------------------------------
\begin{figure}[!htbp]
\begin{minipage}{0.47\textwidth}
 \centerline{\includegraphics[width=1.0\textwidth]{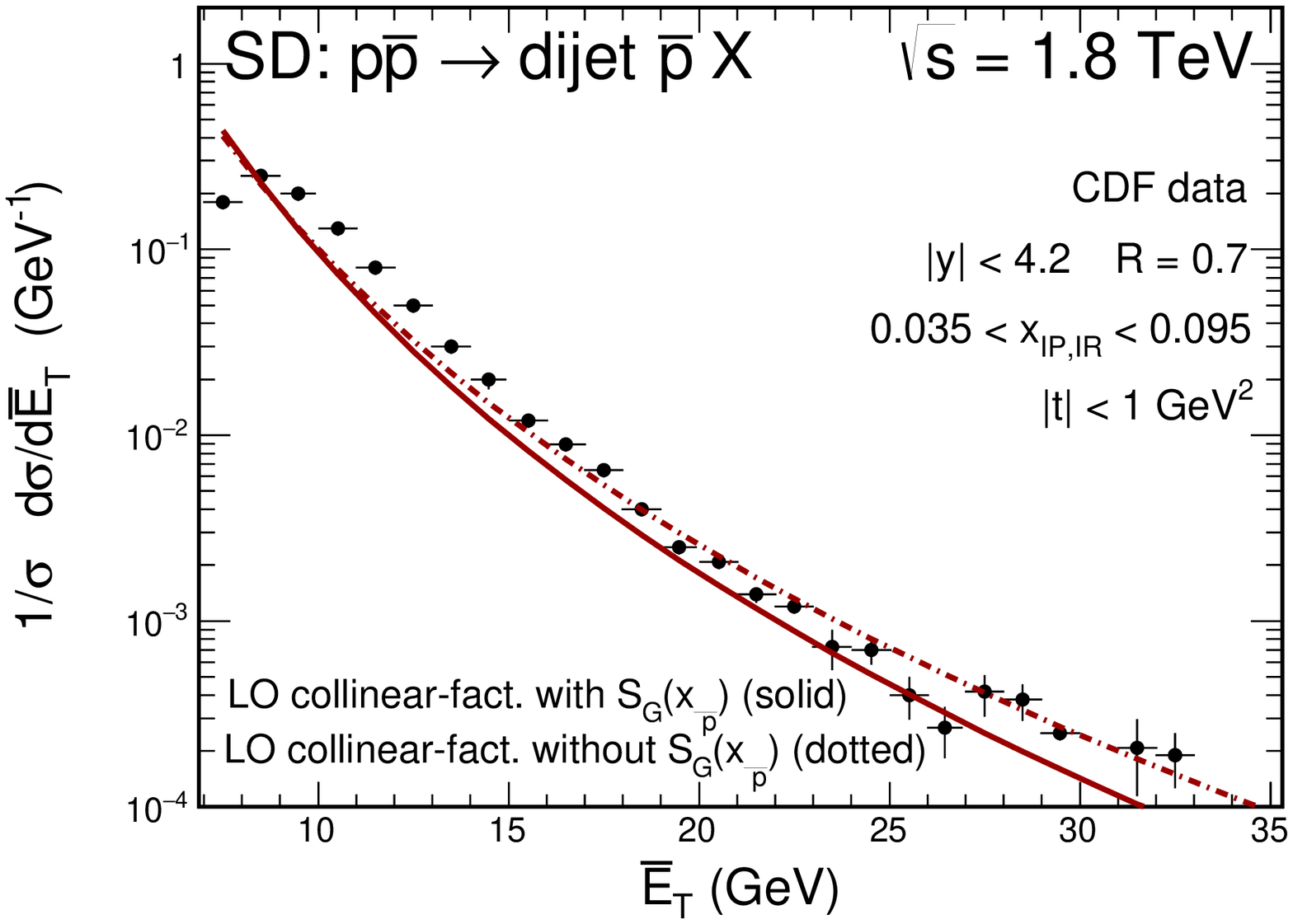}}
\end{minipage}
\hspace{0.5cm}
\begin{minipage}{0.47\textwidth}
 \centerline{\includegraphics[width=1.0\textwidth]{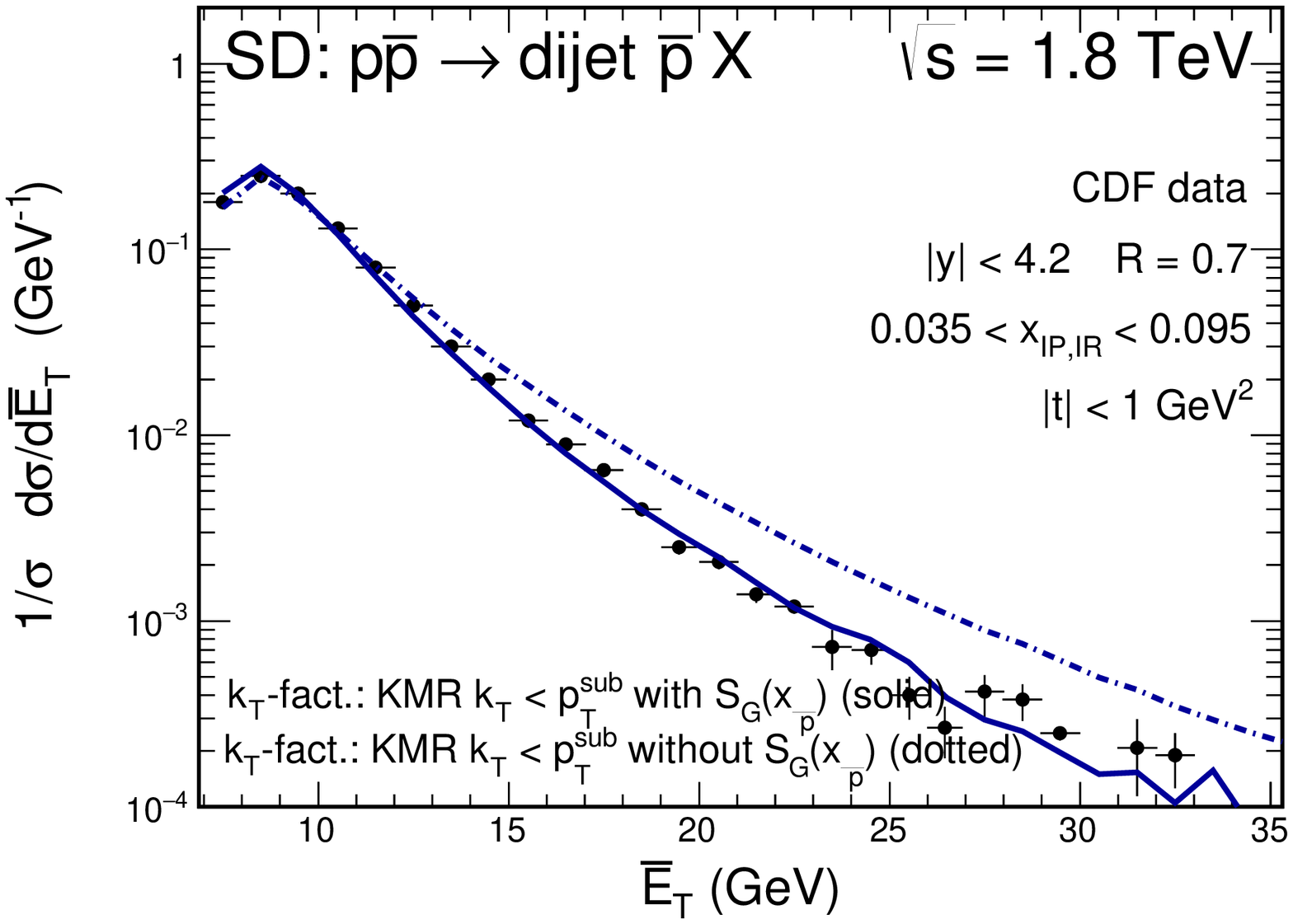}}
\end{minipage}
   \caption{
\small ${\overline E}_T$ distribution for collinear (left panel)
and $k_t$-factorization (right panel) approaches with and without
inclusion of the dependence of $S_G$ on $x_{\bar p}$.
}
 \label{fig:ratio_pt}
\end{figure}

%-----------------------------------------------------------------------------
In Fig.~\ref{fig:ratio_pt} we again show distribution in ${\overline E}_T$
for collinear (left panel) and $k_t$-factorization (right panel) 
approaches. The inclusion of the dependence of $S_G$ on 
$x_{\bar p}$ improves the overall agreement with the CDF data. 

%-----------------------------------------------------------------------------
\begin{figure}[!htbp]
\begin{minipage}{0.47\textwidth}
 \centerline{\includegraphics[width=1.0\textwidth]{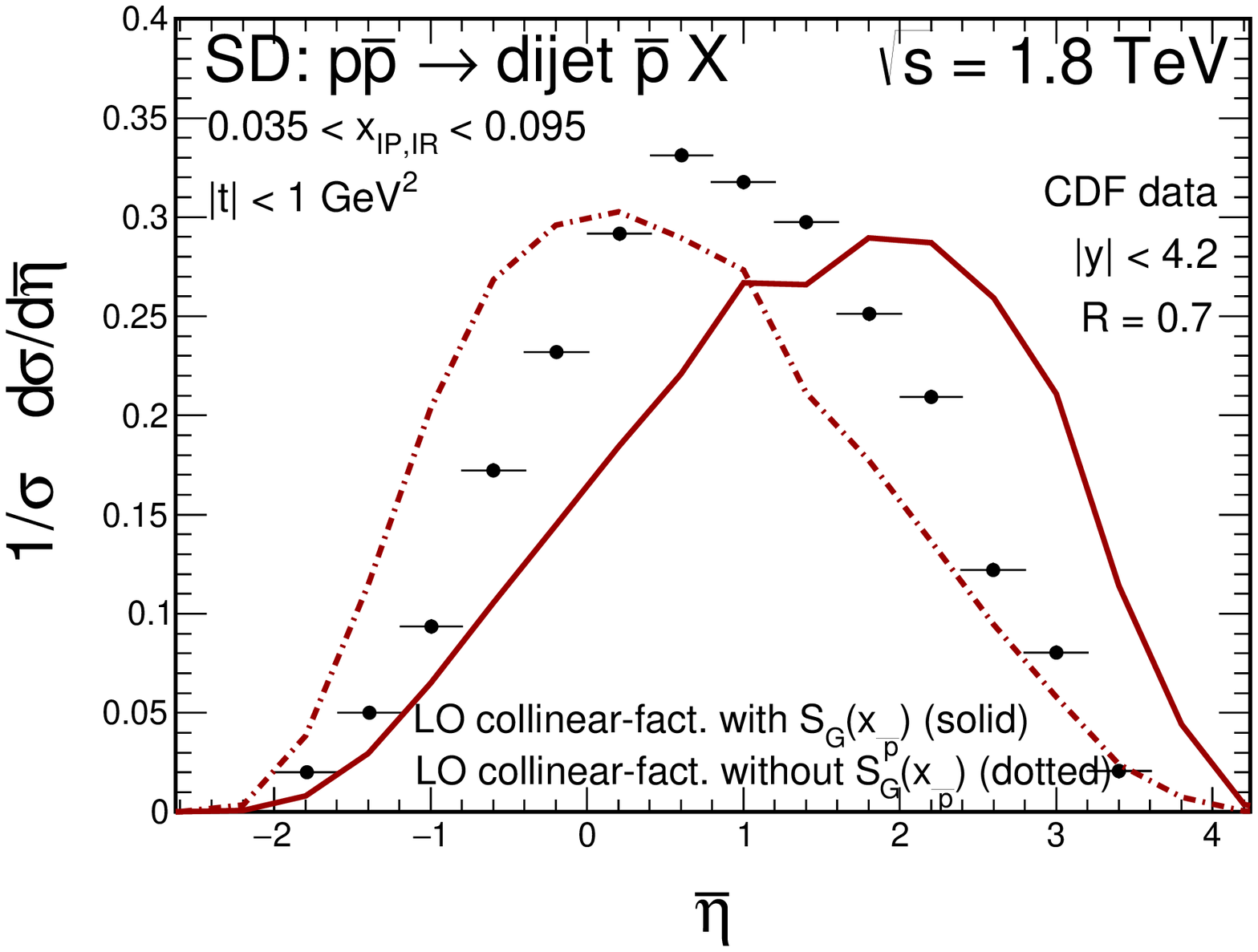}}
\end{minipage}
\hspace{0.5cm}
\begin{minipage}{0.47\textwidth}
 \centerline{\includegraphics[width=1.0\textwidth]{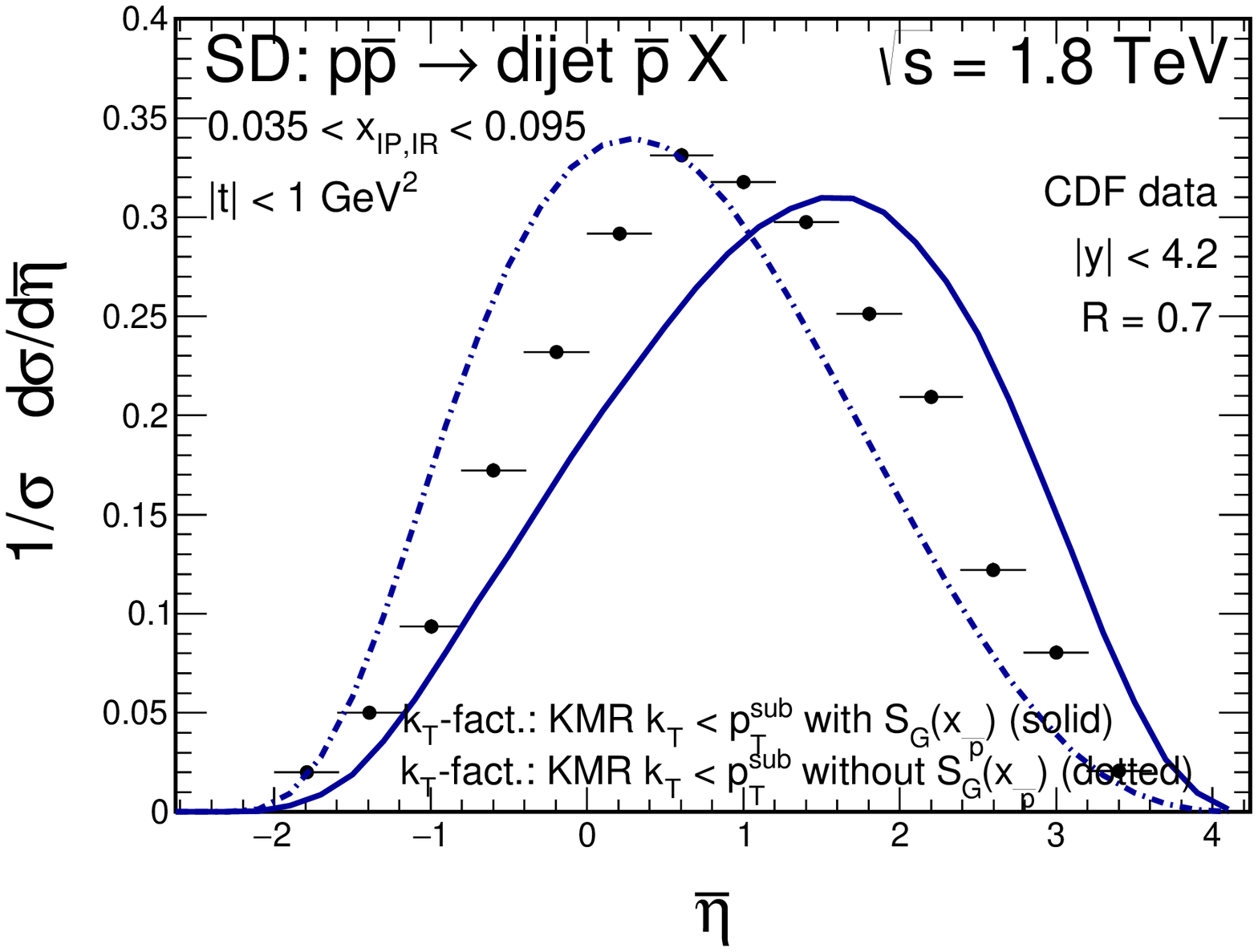}}
\end{minipage}
   \caption{
\small ${\overline \eta}$ distribution for collinear (left panel)
and $k_t$-factorization (right panel) approaches with and without
inclusion of the dependence of $S_G$ on $x_{\bar p}$.
}
 \label{fig:ratio_eta}
\end{figure}

%-----------------------------------------------------------------------------
In Fig.~\ref{fig:ratio_eta} we show similar distributions in ${\overline \eta}$.
One can observe a sizable shift of the distributions towards
larger ${\overline \eta}$. The shift is in a correct direction 
but is much too big. This should be traced back to the extreme 
assumption of the dependence of $S_G$ on $x_{\bar p}$ only. 
In reality $S_G$ may depend on a few kinematical variables. 
However, such a study goes far beyond the scope of the present paper.

%-----------------------------------------------------------------------------
%-----------------------------------------------------------------------------

%-------------------------------------------
\subsection{Predictions for the LHC}
%--------------------------------------------

In this subsection we would like to present our results for the LHC
energy $\sqrt{s}$ = 13 TeV. In our calculations we use cuts relevant
for the planned ATLAS experiments, so we use range of rapidities
relevant for the ATLAS experiment $-4.9 < y_{1}, y_{2} < 4.9$.
We consider rather low cut on the transverse momenta of jets
$p_{t} >$ 20 GeV. In the following we shall use $S_G$ = 0.05.

\begin{figure}[!htbp]
\begin{minipage}{0.47\textwidth}
 \centerline{\includegraphics[width=1.0\textwidth]{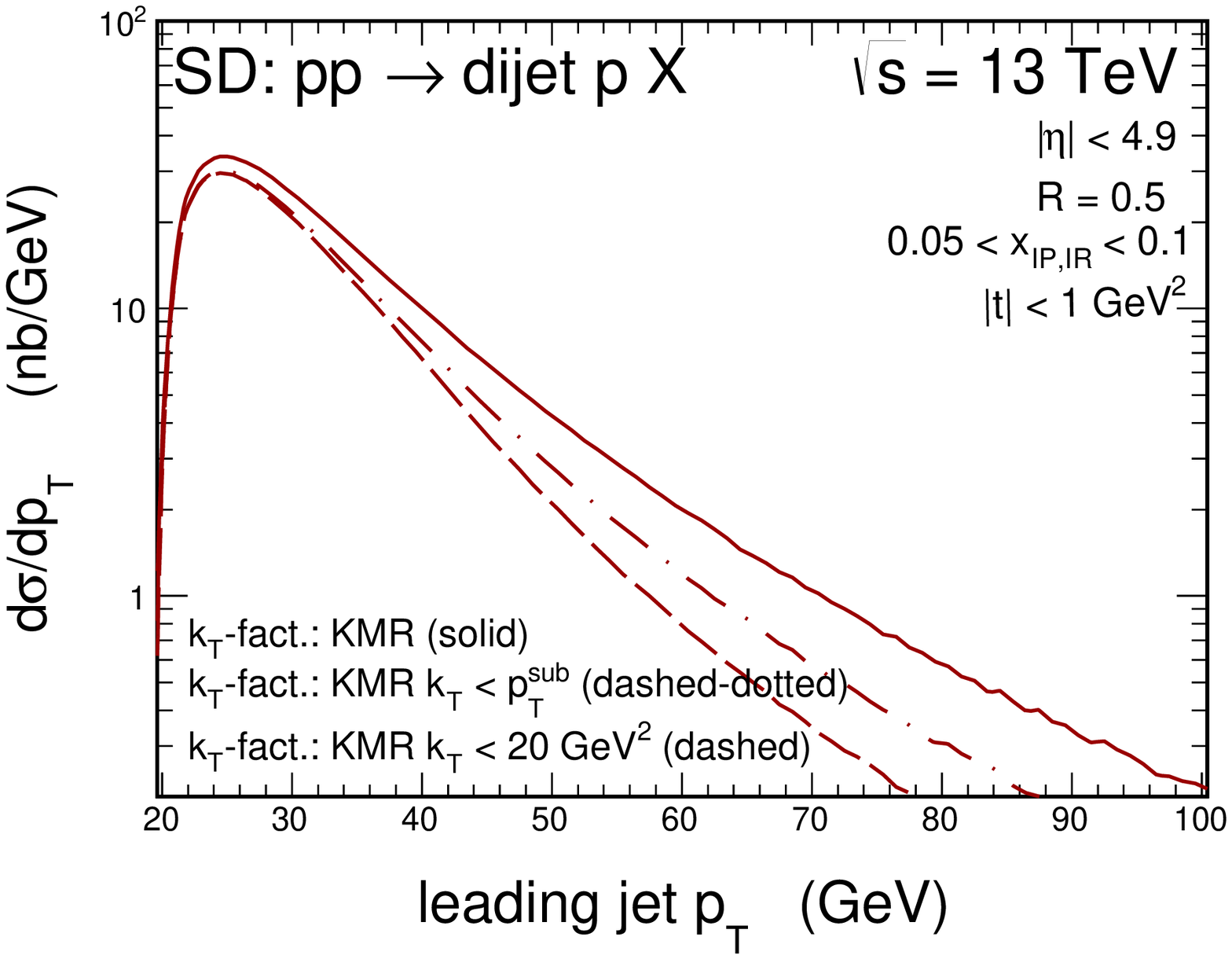}}
\end{minipage}
\hspace{0.5cm}
\begin{minipage}{0.47\textwidth}
 \centerline{\includegraphics[width=1.0\textwidth]{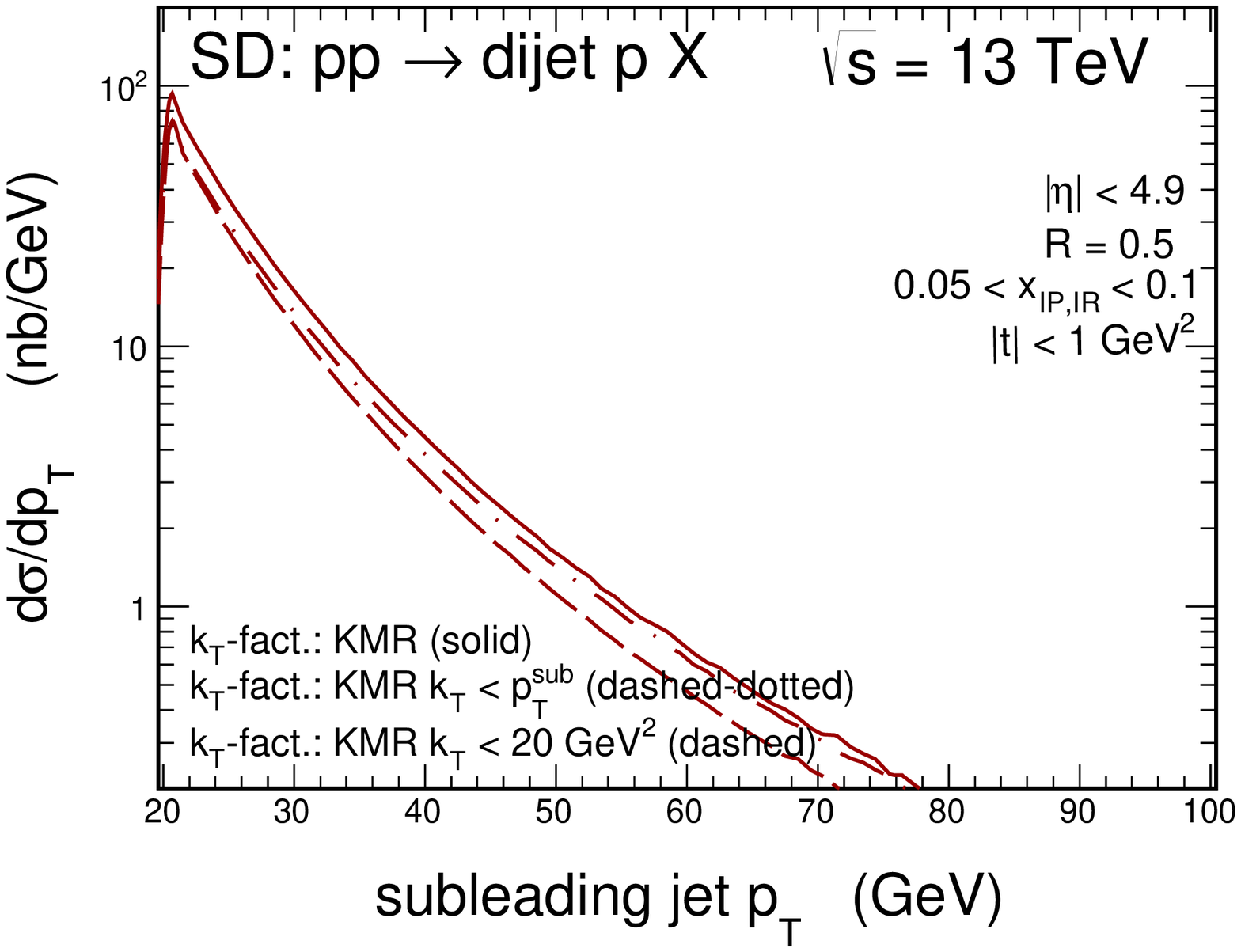}}
\end{minipage}
   \caption{
\small Distribution in the jet transverse momentum for leading
(left panel) and subleading (right panel) for $\sqrt{s}$ = 13 TeV
and for the ATLAS cuts. Here $S_G$ = 0.05.
}
 \label{fig:7}
\end{figure}

In Fig.~\ref{fig:7} we show distribution in jet transverse momentum,
for leading (left panel) and subjeading (right panel) jets.
As for the Tevatron we discuss the role of extra cuts on parton 
transverse momenta. The cuts have bigger efect on leading jets.

%-----------------------------------------------------------------------------
%------------------------------------------------------------------------------

\begin{figure}[!htbp]
\begin{minipage}{0.47\textwidth}
 \centerline{\includegraphics[width=1.0\textwidth]{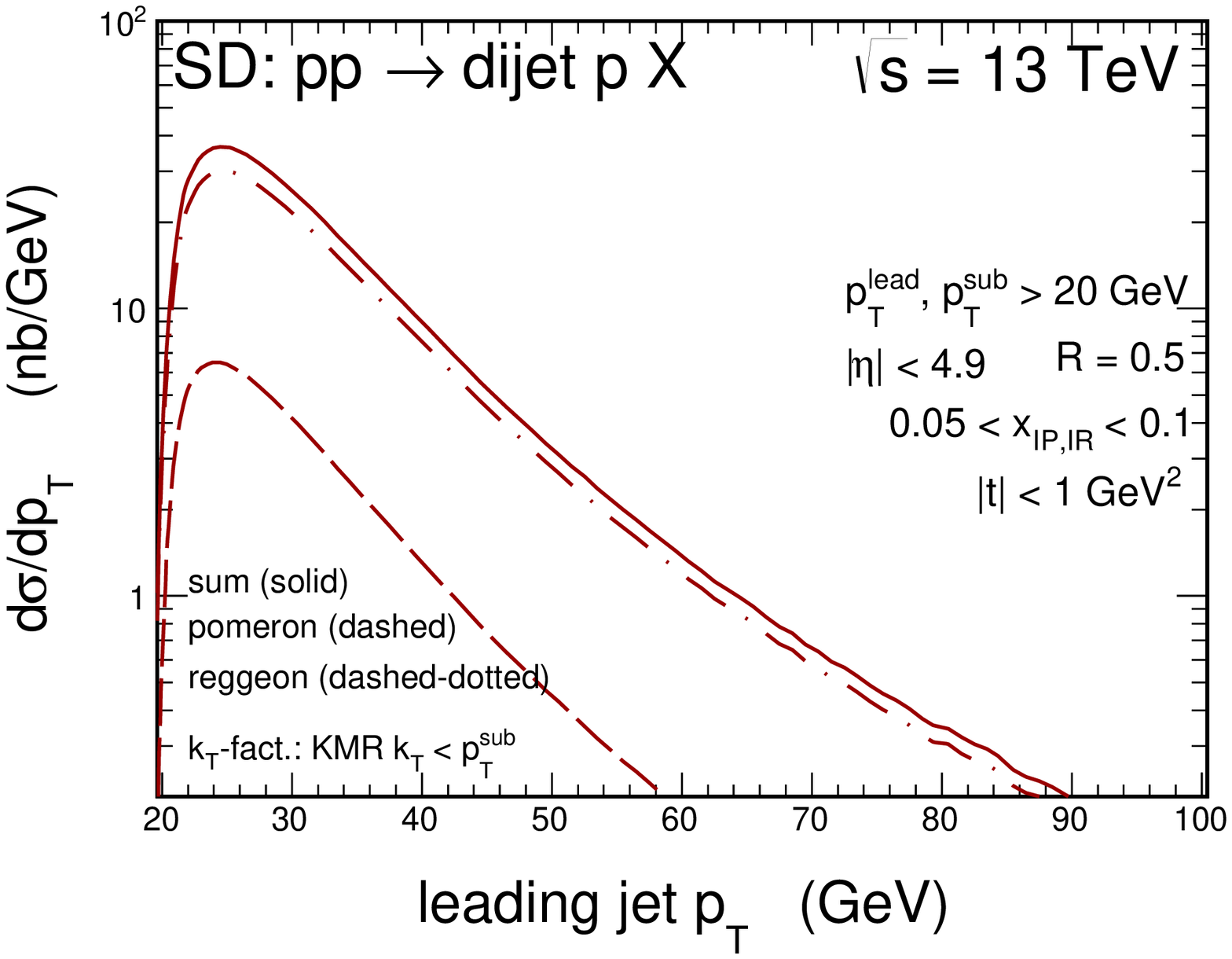}}
\end{minipage}
\hspace{0.5cm}
\begin{minipage}{0.47\textwidth}
 \centerline{\includegraphics[width=1.0\textwidth]{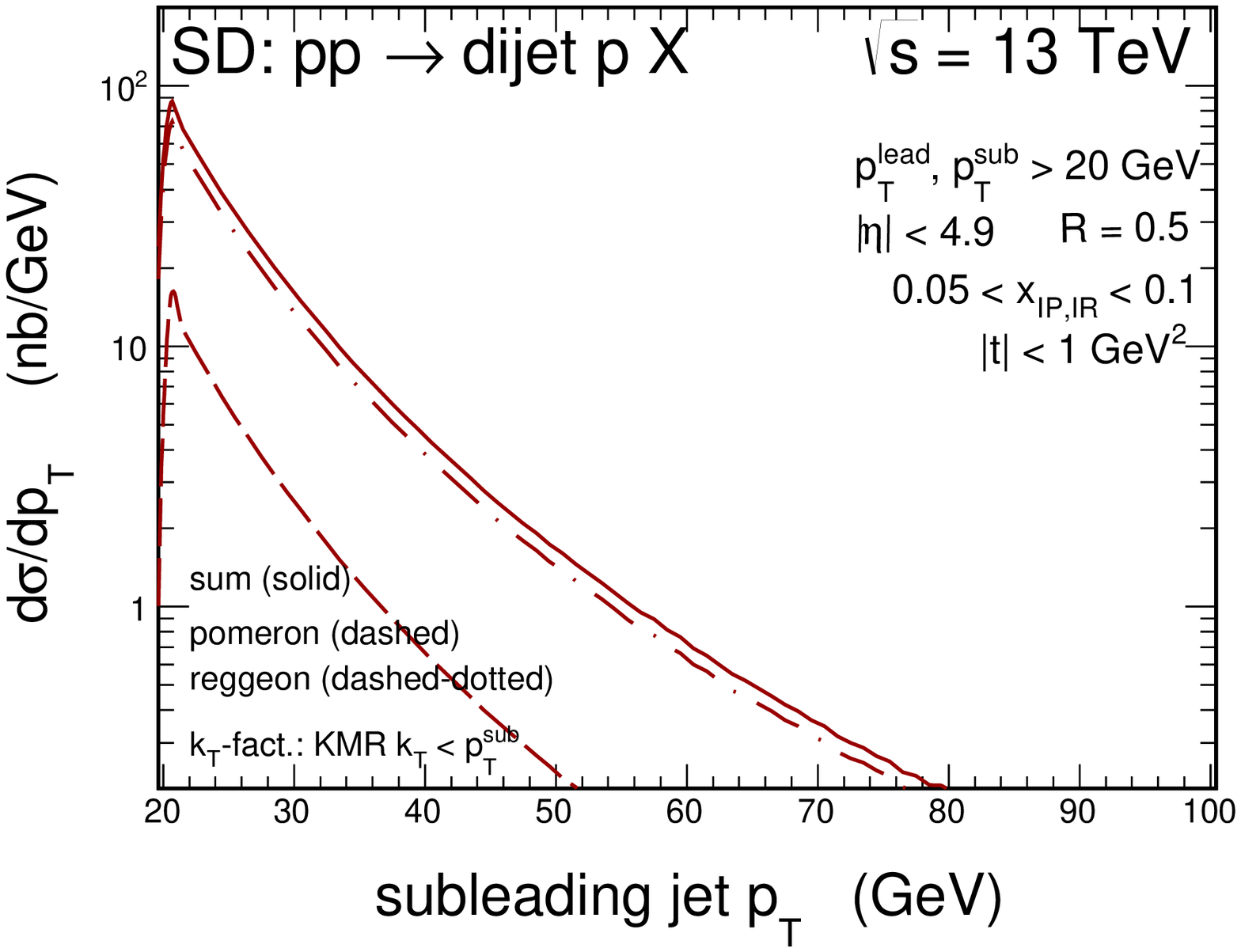}}
\end{minipage}
   \caption{
\small The contribution of pomeron and subleading reggeon for transverse
momentum distribution for leading (left panel) and subleading
(right panel) jet for $\sqrt{s}$ = 13 TeV and for the ATLAS cuts. Here $S_G$ = 0.05.
}
 \label{fig:8}
\end{figure}

In Fig.~\ref{fig:8} we compare contributions of the pomeron and subleading reggeon 
for the ATLAS range of  $x_{I\!P,I\!R}$ . The subleading contribution is larger than 10 \%. There is no evident dependence
on the value of the transverse momentum.

%-----------------------------------------------------------------------------
%------------------------------------------------------------------------------

\begin{figure}[!htbp]
\begin{minipage}{0.47\textwidth}
 \centerline{\includegraphics[width=1.0\textwidth]{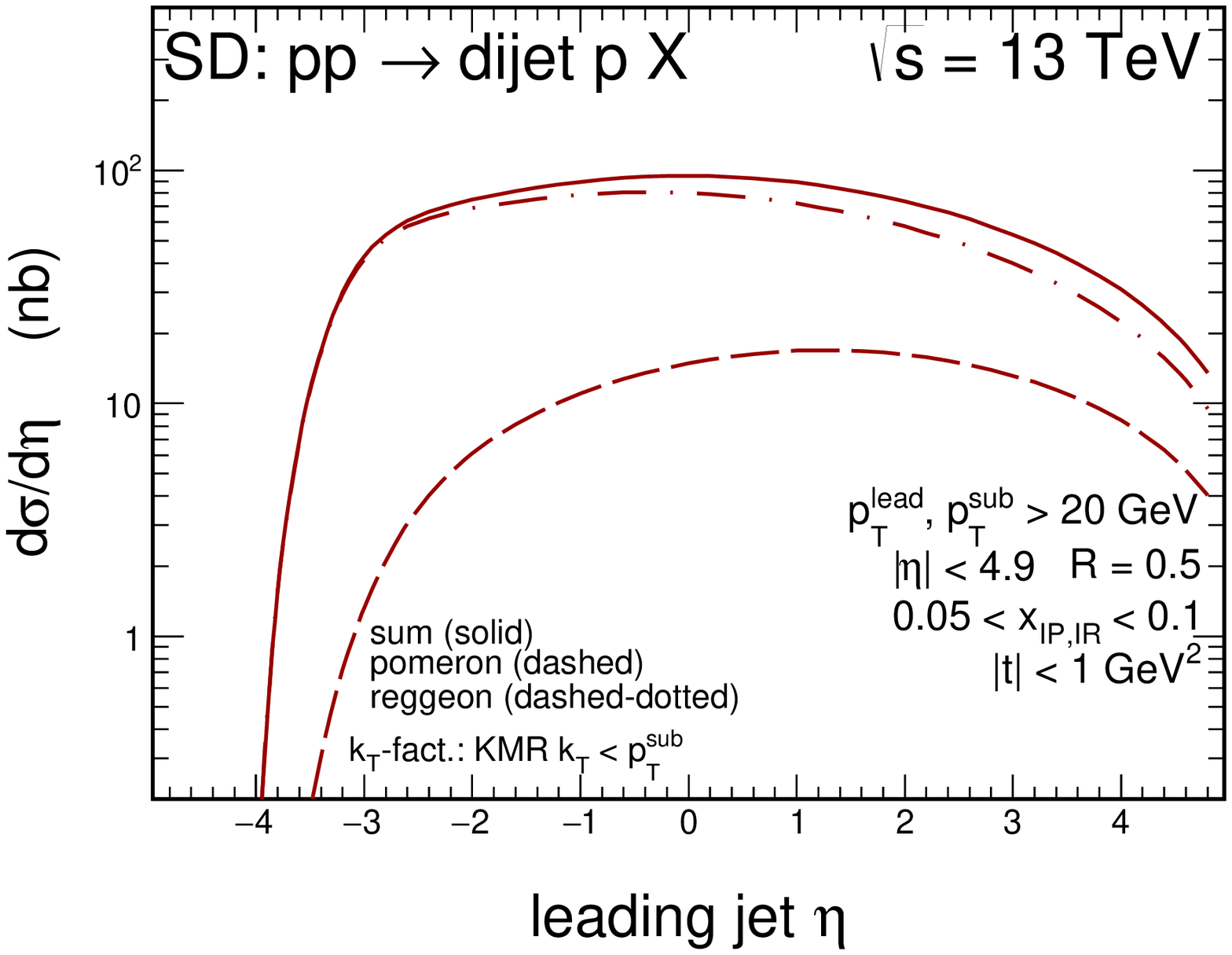}}
\end{minipage}
\hspace{0.5cm}
\begin{minipage}{0.47\textwidth}
 \centerline{\includegraphics[width=1.0\textwidth]{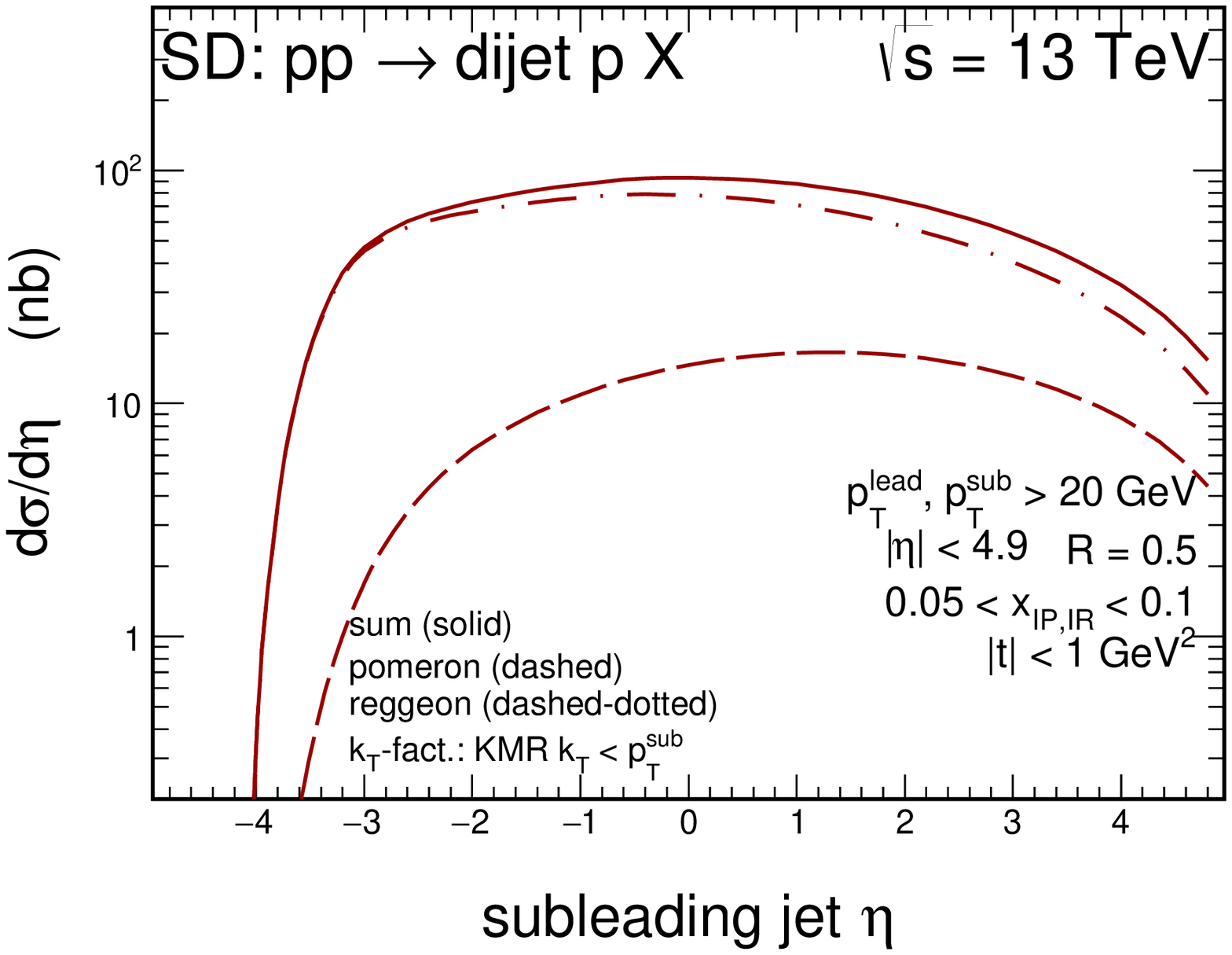}}
\end{minipage}
   \caption{
\small Distribution in the jet rapidity for leading
(left panel) and subleading (right jet) jet for $\sqrt{s}$ = 13 TeV. Here $S_G$ = 0.05.
}
 \label{fig:9}
\end{figure}

In Fig.~\ref{fig:9} we show similar distributions for jet rapidity again
for leading and subleading jet. As previously we show contributions
of pomeron and subleading reggeon separately.
Here the relative contribution of subleading reggeon is an evident 
function of rapidity, both for leading and subleading jet.

%-----------------------------------------------------------------------------
\begin{figure}[!htbp]
\begin{minipage}{0.47\textwidth}
 \centerline{\includegraphics[width=1.0\textwidth]{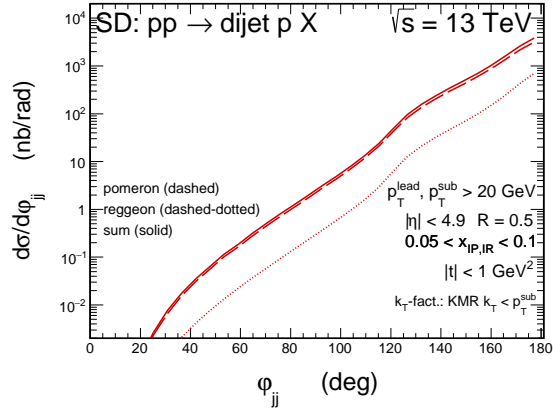}}
\end{minipage}
   \caption{
\small Our predictions for azimuthal angle correlations between leading and
subleading jet. Here $S_G$ = 0.05.
}
 \label{fig:10}
\end{figure}

%------------------------------------------------------------------------------
Azimuthal angle correlations between the leading and subleading jet
are shown in Fig.~\ref{fig:10}. Similar shapes are obtained for pomeron and reggeon contributions.

%-----------------------------------------------------------------------------
%------------------------------------------------------------------------------

\begin{figure}[!htbp]
\begin{minipage}{0.47\textwidth}
 \centerline{\includegraphics[width=1.0\textwidth]{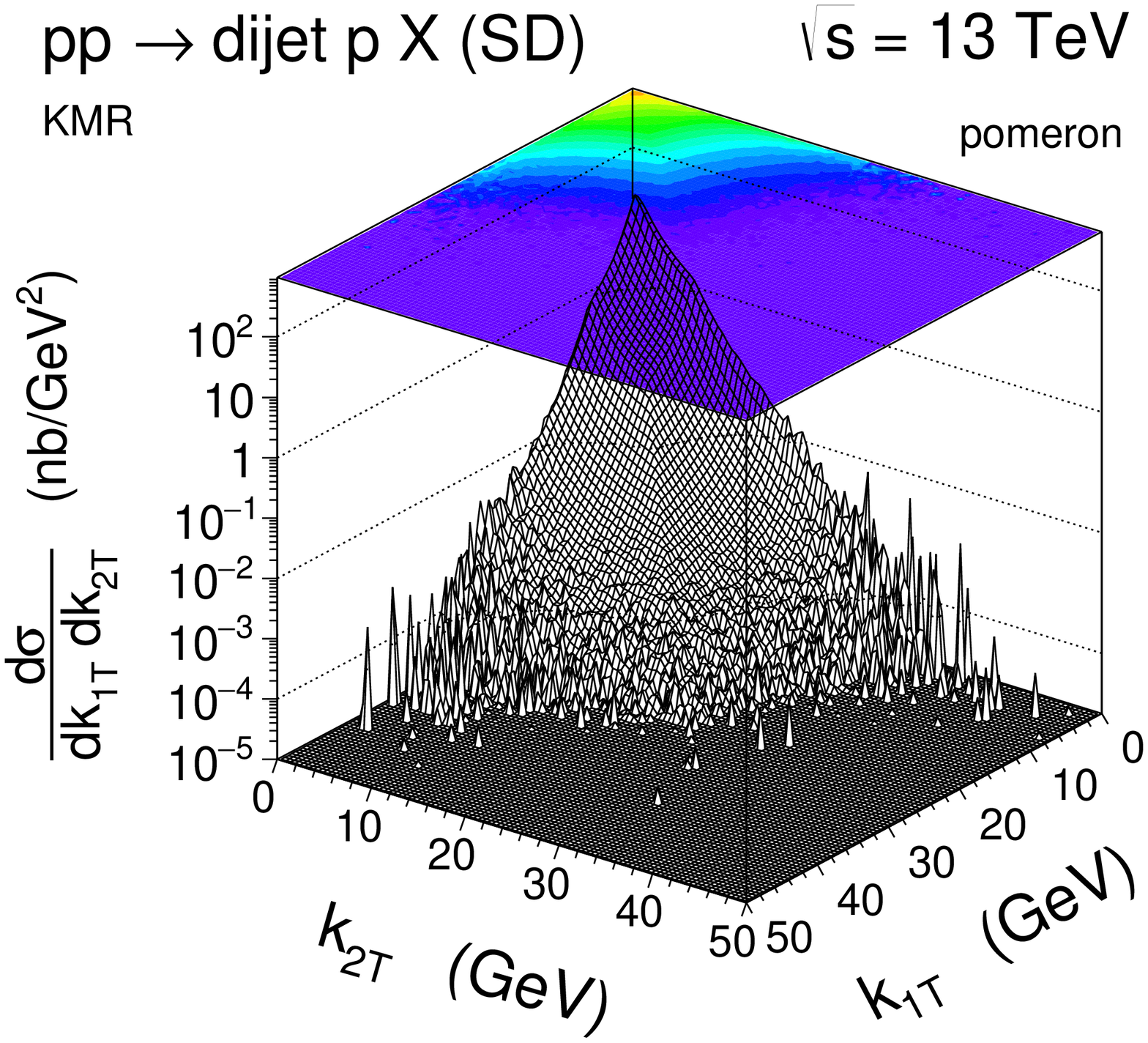}}
\end{minipage}
\hspace{0.5cm}
\begin{minipage}{0.47\textwidth}
 \centerline{\includegraphics[width=1.0\textwidth]{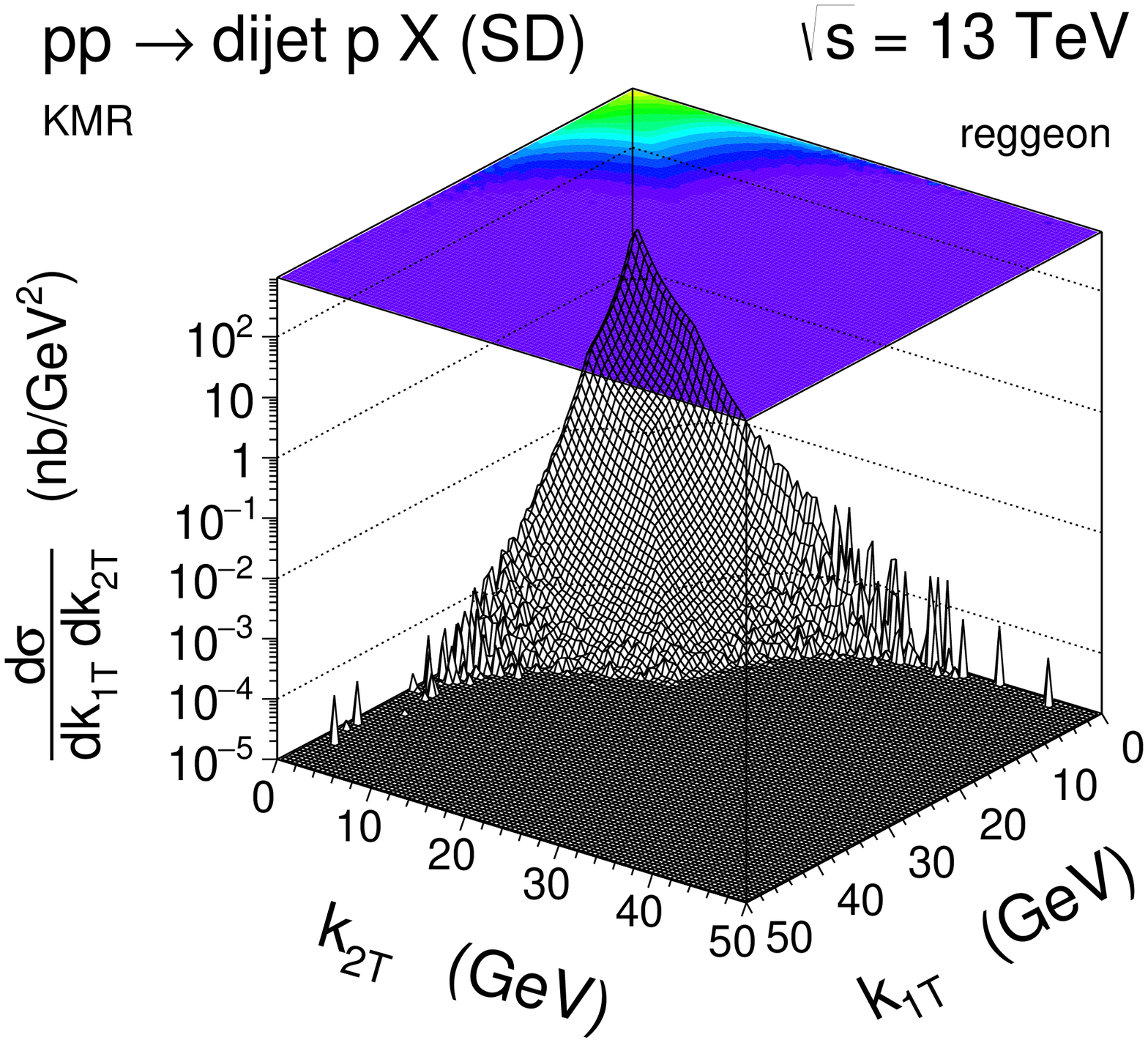}}
\end{minipage}
   \caption{
\small Two-dimensional distributions in parton transverse momenta
for pomeron (left panel) and subleading reggeon (right panel).
In this calculation $\sqrt{s}$ = 13 TeV and ATLAS cuts were imposed. Here $S_G$ = 0.05.
}
 \label{fig:11}
\end{figure}

Finally in Fig.~\ref{fig:11} we show purely theoretical two-dimensional 
distributions in transverse momenta of partons for pomeron 
(left panel) and subleading reggeon (right panel), respectively,
for nondiffractive and diffractive sides.
As for the Tevatron the distributions are surprisingly symmetric 
in $k_{1T}$ and $k_{2T}$. In this calculation no extra cuts on 
parton transverse momenta have been imposed. 
We stress that very large transverse momenta of partons enter
the considered dijet production.

%-----------------------------------------------------------------------------
%------------------------------------------------------------------------------

In Table I we present the integrated cross section for the ATLAS acceptance
for single-diffractive production of dijets for different cuts of the jet-$p_{T}$.

%------------------------------------------------------------------------------------------------------------------------------
\begin{table}[tb]%
\caption{The calculated cross sections in microbarns for single-diffractive production of dijets in $pp$-scattering at $\sqrt{s} =$ 13 TeV for different cuts on transverse momentum of the dijets. 
Here, the rapidity of the dijets is $|y^{jet}| < 4.9$, that corresponds to the ATLAS detector acceptance. The cross section here is not multiplied by the gap survival factors.}

\label{tab:cross sections}
\centering %
%\newcolumntype{Z}{>{\centering\arraybackslash}X}
%\newcommand{\tn}{\tabularnewline}
\resizebox{\textwidth}{!}{%
\begin{tabularx}{0.9\linewidth}{c c c c c}
\\[-4.ex] 
\toprule[0.1em] %
\\[-4.ex] 
%\\[1.0ex]

\multirow{2}{3.cm}{$p_{T,min}^{jet}$ cuts}     & \multirow{1}{3.5cm}{collinear} & \multicolumn{3}{c}{$k_{T}$-factorization approach}   \\ [-0.2ex]
                      & \multirow{1}{3.5cm}{{\small MMHT2014nlo}}    &     \multirow{1}{1.5cm}{KMR}         &   \multirow{1}{3.5cm}{{\scriptsize KMR $k_{T} < p_{T,min}^{jet}$ (IP)}}     &   \multirow{1}{3.5cm}{{\scriptsize KMR $k_{T} < p_{T,min}^{jet}$ (IR)}}  \\ [-0.2ex]
\bottomrule[0.1em]

\multirow{1}{3.cm}{$p_{T}^{jet} > 20$ GeV}  & \multirow{1}{3.5cm}{$\qquad 9.08$} & \multirow{1}{1.5cm}{11.42} & \multirow{1}{3.5cm}{$\qquad 8.53$} & \multirow{1}{3.0cm}{$\quad 1.79$} \\  [-0.2ex]
\multirow{1}{3.cm}{$p_{T}^{jet} > 35$ GeV}  & \multirow{1}{3.5cm}{$\qquad 2.34$} & \multirow{1}{1.5cm}{3.89} & \multirow{1}{3.5cm}{$\qquad 3.89$} & \multirow{1}{3.0cm}{$\quad 0.62$} \\  [-0.2ex]
\multirow{1}{3.cm}{$p_{T}^{jet} > 50$ GeV}  & \multirow{1}{3.5cm}{$\qquad 0.42$} & \multirow{1}{1.5cm}{0.83} & \multirow{1}{3.5cm}{$\qquad 0.68$} & \multirow{1}{3.0cm}{$\quad 0.16$} \\  [-0.2ex]

\hline

\bottomrule[0.1em]

\end{tabularx}
}
\end{table}
%--------------------------------------------------------------------------------

%--------------------------
\section{Conclusions}
%--------------------------

In the present paper we have presented for the first time results for
the single-diffractive production of dijets within $k_t$-factorization
approach.  The resolved pomeron model with flux of pomeron and reggeon 
and parton distribution in pomeron have been used.
The diffractive unintegrated parton distributions were obtained based 
on their collinear counterparts.
The latter were used to fit the HERA data for diffractive $F_2$
structure function and for diffractive dijet production.
The rapidity gap is not calculated but can be fitted to the data.  
A constant value was assumed as a default.

Results of our calculations were compared with the Tevatron data 
where forward antiprotons and rapidity gaps were measured. We have 
calculated distributions in ${\overline{E}_T}$ and ${\overline \eta}$. 
A resonable agreement has been achieved.
We have compared results obtained within collinear and 
$k_t$-factorization approaches.
The $k_t$-factorization leads to a better description in 
$E_T$ close to the lower transverse momentum cut.

Several other distributions have been presented and discussed, many 
of them for a first time.

It is rather difficult to describe the distributions in 
$x_{\overline p}$ with a constant value of gap survival factor, 
especially for $\sqrt{s}$ = 1.8 TeV.
We have considered a possibility that the gap survival factor 
depends exclusively on $x_{\overline p}$ and studied consequences 
for other observables. A phenomenological $x_{\overline p}$ function
was used to fit the Tevatron data.  Such a dependence of the gap 
survival factor leads to an effective shift of the distribution in 
${\overline  \eta}$ in better agreement with the Tevatron data.
Our preliminary study suggest that the dependence of gap survival factor
on kinematical variables can be also an important ingredient in order 
to understand details of rapidity distributions.
Clearly further studies are necessary in a future.

We have  also made predictions for future LHC measurements.  
Several differential distributions have been presented.  
We hope for their verification in a near future.

\vspace{1cm}

{\bf Acknowledgements}

This study was partially supported by the Polish National Science Centre grants DEC-2013/09/D/ST2/03724 and DEC-2014/15/B/ST2/02528.

%-------------------------------------------------------------------------------------

\end{document}